\documentclass[twocolumn]{aastex63}
\turnoffedit
\usepackage{amsmath,amssymb,amstext,fp,mathtools,xspace}

\newcommand{\Cov}{\operatorname{Cov}}

 \newcommand{\f}[1]{\mathrm{\uppercase{f{#1}}}}
 \let\F=\f

 \newcommand{\logg}{\log (g)}
 \newcommand{\Teff}{T_{\mathrm{eff}}}
 \let\teff=\Teff

 \newcommand{\hst}{\emph{HST}\xspace}
 \let\HST=\hst
 
 \let\JWST=\jwst

 \let\roman=\rst

 \makeatletter
 \renewcommand{\sectionautorefname}{\S\@gobble}
 \renewcommand{\subsectionautorefname}{\S\@gobble}
 \renewcommand{\subsubsectionautorefname}{\S\@gobble}
 \makeatother

\begin{document}

\title{Empirical 2MASS-WFC3/IR filter transformations across the HR diagram from synthetic photometry}
\shorttitle{NIR filter transformations}
\shortauthors{M.~J.~Durbin et al.}

\author[0000-0001-7531-9815]{M.~J.~Durbin}
\affiliation{Department of Astronomy, University of California, Berkeley, Berkeley, CA, 94720, USA}
\affiliation{Department of Astronomy, University of Washington, Box 351580, U.W., Seattle, WA 98195-1580, USA}

\author[0000-0002-1691-8217]{R.~L.~Beaton}
    \affil{Space Telescope Science Institute, Baltimore, MD, 21218, USA}
    \affil{Department of Physics and Astronomy, Johns Hopkins University, Baltimore, MD 21218, USA}
    \affiliation{Department of Astrophysical Sciences, Princeton University, 4 Ivy Lane, Princeton, NJ 08544, USA}
    \affiliation{The Observatories of the Carnegie Institution for Science, 813 Santa Barbara St., Pasadena, CA~91101}

\author[0000-0002-0048-2586]{A.~J.~Monson}
    \affiliation{Department of Astronomy \& Astrophysics, The Pennsylvania State University, 525 Davey Lab, University Park, PA 16802, USA}
    \affiliation{Department of Astronomy/Steward Observatory, University of Arizona, 933 North Cherry Avenue, Tucson, AZ 85721, USA}

\author[0000-0002-3749-4978]{B.~Swidler}
\affiliation{Department of Astrophysical Sciences, Princeton University, 4 Ivy Lane, Princeton, NJ 08544, USA}

\author[0000-0002-1264-2006]{J.~J.~Dalcanton}
    \affiliation{Center for Computational Astrophysics, Flatiron Institute, 162 Fifth Avenue, New York, NY 10010, USA}
    \affiliation{Department of Astronomy, University of Washington, Box 351580, U.W., Seattle, WA 98195-1580, USA}

\begin{abstract}

Near-infrared bandpasses on spaceborne observatories diverge from their ground-based counterparts as they are free of atmospheric telluric absorption. 
Available transformations between respective filter systems in the literature rely on theoretical stellar atmospheres, which are known to have difficulties reproducing observed spectral energy distributions of cool giants. 
We present new transformations between the 2MASS $JHK_S$ and \HST WFC3/IR F110W, F125W, \& F160W photometric systems based on synthetic photometry of empirical stellar spectra from four spectral libraries. 
This sample comprises over 1000 individual stars, which together span nearly the full HR diagram and sample stellar populations from the solar neighborhood out to the Magellanic Clouds, covering a broad range of ages, metallicities, and other relevant stellar properties. 
In addition to global color-dependent transformations, we examine band-to-band differences for cool, luminous giant stars in particular, including multiple types of primary distance indicators.
\end{abstract}


%%%%%%%%%%%%%%%%%%%%%%%%%%%%%%%%%%%%%%%%%%%%%%%%%%%%%%%%%%%%%%%%%%%%%%%%%%%%%%
\section{Introduction} \label{sec:intro}
%%%%%%%%%%%%%%%%%%%%%%%%%%%%%%%%%%%%%%%%%%%%%%%%%%%%%%%%%%%%%%%%%%%%%%%%%%%%%%
Near-infrared (NIR) spectrophotometry of stars is increasingly important to a wide range of astrophysical issues.
Recent and upcoming missions with NIR capabilities, such as \JWST and \roman, will resolve orders of magnitude more stars at higher precision and greater distances than their closest predecessors.

In practice, however, the potential of these and other space-based observations is frequently limited by the ability to link new photometry to the rich legacy of ground-based observations \citep[e.g. 2MASS, ][]{2006AJ....131.1163S}. 
The widely-used $JHK$ filter set and its near relatives are largely shaped to accommodate the atmospheric windows where the Earth's atmosphere is transparent.
Space-based instruments, however, do not share these constraints, and can sample the full range of the stellar spectrum. 
Even for space-based filters expressly designed to mimic their ground-based counterparts, subtle differences can arise from the atmospheric absorption that typically defines ground filters' blue and red edges.
Similarly, spatial and temporal variations in Earth's atmosphere add complexity to the task of consistently calibrating ground-based NIR photometry, both night-to-night and across observatories. 

\begin{figure}
    \centering
    \includegraphics[width=\columnwidth]{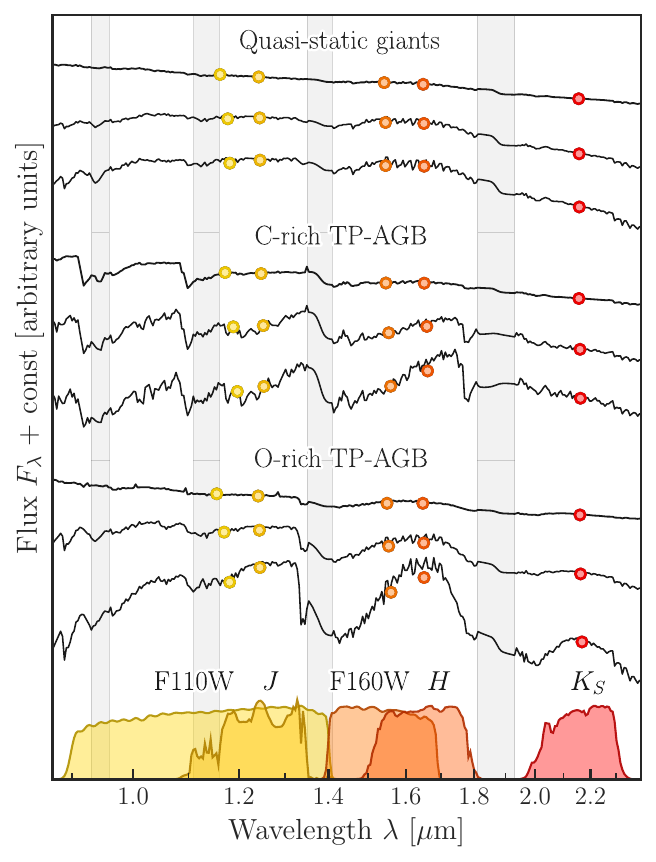}
    \caption{Schematic comparison of near-infrared stellar spectra \citep[black lines,][]{2022A&A...661A..50V} and their integrated \edit1{flux densities} (colored points, plotted at respective effective wavelengths) for three broad classes of cool and luminous giants.
    These include non-pulsating AGB and upper RGB stars (upper three), and thermally pulsating carbon-rich (center) and oxygen-rich (lower) AGB stars respectively.
    Transmission curves for the 2MASS $JHK_S$ and \HST F110W, F160W bandpasses are shown as filled curves at the bottom, and regions with significant atmospheric telluric absorption are marked by gray vertical bands.
    }
    \label{fig:filter_curves}
\end{figure}

Interpreting even precisely calibrated NIR photometry is particularly challenging for the cool and luminous giants that dominate the integrated rest-frame infrared light of intermediate-aged to ancient stellar populations.
In such stars, low surface gravities ($\log g \lesssim 1$) and effective temperatures ($\Teff \lesssim 4000$~K) give rise to deep and wide atomic and molecular absorption features that can have profound impacts on spectral shapes as measured with typical broadband filters.
\autoref{fig:filter_curves} shows example spectral sequences for three broad subclasses of such stars: non-pulsating asymptotic giant branch (AGB) or bright first-ascent red giant branch (RGB) stars \citep[``quasi-static'' giants, following][]{2022A&A...661A..50V}, and thermally pulsing AGB (TP-AGB) stars with carbon- and oxygen-rich atmospheres.
We overlay synthetic integrated \edit1{flux densities} for each spectrum in the ground-based 2MASS $JHK_S$ and {\emph{Hubble Space Telescope}} \edit1{Wide Field Camera 3} (\HST\edit1{/WFC3}) F110W and F160W filters.
While the integrated \edit1{flux densities} in overlapping \HST and 2MASS bandpasses approximately agree for the quasi-static giants, they begin to diverge as molecular feature strength increases in both TP-AGB subtypes.
Furthermore, even non-pulsating giants exhibit variability over a range of amplitudes, timescales, and modes due to asteroseismic oscillations, further complicating their photometric calibration.

In this paper, we begin a systemic empirical cross-calibration of the ground-based 2MASS filter set with a subset of popular broadband filters on the the infrared channel of \HST's Wide Field Camera 3 (WFC3/IR).
We take advantage of several empirical spectral libraries that have been released in the past two decades, which together include stars ranging in temperature from below 3000~K to over 60,000~K.
At the higher temperatures, the sample is dominated by main sequence stars and white dwarfs, and at cool temperatures, the sample is approximately evenly split between low-mass dwarfs and evolved giants.

\autoref{sec:obs} describes the data products used in this work, including all spectral libraries, distance and extinction estimates, and other literature data used in determining and interpreting our final synthetic magnitudes and colors.
\autoref{sec:analysis} describes our synthetic photometry production and validation process.
In \autoref{sec:res} we present best-fit global 2MASS$\leftrightarrow$\HST/WFC3-IR magnitude and color transformations.
We compare to previously published relations and examine residuals with respect to stellar properties in \autoref{sec:dis}, and discuss conclusions and future work in \autoref{sec:conc}.

%%%%%%%%%%%%%%%%%%%%%%%%%%%%%%%%%%%%%%%%%%%%%%%%%%%%%%%%%%%%%%%%%%%%%%%%%%%%%%
\section{Data} \label{sec:obs}
%%%%%%%%%%%%%%%%%%%%%%%%%%%%%%%%%%%%%%%%%%%%%%%%%%%%%%%%%%%%%%%%%%%%%%%%%%%%%%
% update table 1 for dr3 + check others

\newcolumntype{Y}[1]{>{\centering\arraybackslash}p{#1}}
\begin{table*}[htbp]
    \centering
    \caption{Summary of spectral library datasets used in this work.} \label{tab:spec_libraries_summary}            
    \begin{tabular}{r|c c c c c c }
    \hline \hline 
        {} & (E)IRTF & XSL DR3 & \multicolumn{3}{c}{CALSPEC} & All \\
    \hline \hline
        Facility            &  IRTF/SPEX            & VLT/X-shooter              & STIS & NICMOS & WFC3/IR & ---\\ 
        Wavelength range [$\micron$] &  0.8--2.5+    &  0.3--2.45                  & 0.1--1.0 & 0.8--2.5 & 0.8--1.7 & ---\\
        Spectral resolution &  $\sim$2000           &  $\sim$10000               & $>$500 & $\sim$200 & $\sim$170 & ---\\ 
        Absolute fluxing?   &  Yes, to 2MASS        & No                         & \multicolumn{3}{c}{Yes} & ---\\
        Primary reference(s)  & {\citet{2009ApJS..185..289R}}   & {\citet{2022AA...660A..34V}} & \multicolumn{3}{c}{\citet{1990AJ.....99.1243T}} & ---\\ 
        {} & {\citet{2017ApJS..230...23V}}   & {} & \multicolumn{3}{c}{\citet{2014PASP..126..711B}} & {} \\ 
    \hline 
        Spectral types      & primarily late-type   & many & \multicolumn{3}{c}{primarily dwarfs} & ---  \\ 
        Total spectra         & 431 & 736 & \multicolumn{3}{c}{102} & 1269 \\
        Unique stars          & 422 & 614 & \multicolumn{3}{c}{102} & 1061 \\
    \hline
        $J-K_S$ color range   & $-0.1 - 2.5$ & $-0.2 - 2.4$ & \multicolumn{3}{c}{$-0.3 - 0.9$} & $-0.3 - 2.5$ \\
    \hline \hline
    \end{tabular}
\end{table*}

In this section, we describe the datasets used in this work: 
    empirical spectral libraries (\autoref{ssec:libraries}); 
    distance and reddening measurements (\autoref{ssec:distance_reddening});
    and other compiled literature parameters (\autoref{ssec:mixed_library}).

\subsection{Empirical Spectral Libraries with NIR Coverage}\label{ssec:libraries}

Empirical spectral libraries are critical benchmarks for understanding many facets of stellar atmospheres, evolution, and populations \citep{2012ASInC...6....1T}.
Many existing libraries offer excellent coverage of optical and near-UV wavelengths \citep[e.g.\@ ][]{2019ApJ...883..175Y, 
2018MNRAS.480.4766W, 2006MNRAS.371..703S, 2011A&A...532A..95F, 2001A&A...369.1048P, 2004astro.ph..9214P, 2007astro.ph..3658P, 2006hstc.conf..209G, 2004ApJS..152..251V, 2003A&A...402..433L, 2003Msngr.114...10B}.
However, there is a relative dearth of comparable libraries available for the near-infrared.
Such observational libraries are essential, given that theoretical spectra are known to have difficulty reproducing observed NIR SEDs and spectral features of cool and luminous giants \citep{2000A&A...353..322A, 2003A&A...412..481T, 2004A&A...415..571B, 2005MNRAS.362..799M, 2006ApJ...645.1102L, 2012A&A...543A..75L, 2016MNRAS.457.3611A, 2017A&A...601A.141G, 2018MNRAS.473.4698B, 2018MNRAS.476.4459D, 2020MNRAS.491.2025C, 2021A&A...649A..97L, 2019IAUS..343...93A, 2022MNRAS.512..378E, 2023A&A...673A..21E}.

The spectral library datasets adopted in this work were selected to have continuous spectral coverage in the range of at least $0.8 \lesssim \lambda \lesssim 1.8~\micron$, with the SED continuum preserved.
In addition to the descriptions to follow, basic characteristics of the libraries and the stars adopted from them are provided in \autoref{tab:spec_libraries_summary}. 
For all of the stars selected from the libraries, we also search the literature and archives such as SIMBAD for relevant optical and infrared photometry, stellar parameters and classifications, and distances and line-of-sight extinctions.
Because this information is extensive and potentially of use for other purposes, we include machine-readable tables of our final compiled database in \autoref{app:tables}.

\subsubsection{CALSPEC} \label{sssec:calspec}

CALSPEC is a library of spectrophotometric standards that forms the basis of the \HST absolute flux calibration scale\edit1{\footnote{\url{https://archive.stsci.edu/hlsps/reference-atlases/cdbs/calspec/}}} 
\citep[][and references therein]{1990AJ.....99.1243T, 2001AJ....122.2118B, 2007ASPC..364..315B, 2008AJ....136.1171B, 2014PASP..126..711B, 2019AJ....157..229B}.
The majority of observed spectra are taken with the Space Telescope Imaging Spectrograph (STIS), the infrared channel of the Wide Field Camera 3 (WFC3/IR), and/or the Near Infrared Camera and Multi-Object Spectrometer (NICMOS) instruments.
% Some Fun Details: 
STIS covers 0.011-1.03~$\micron$ at low to medium resolution with a number of gratings/grisms; only the G750L and G750M span into the IR ($\sim$0.5- 1.~$\micron$) with resolutions of $\sim$500-1000 and $\sim5000-9000$, respectively.
The WFC3/IR grisms G102 and G140 together span 0.8-1.7~$\micron$, at resolutions of $R=210$ at 1~$\micron$ and 130 at 1.4~$\micron$. 
The NICMOS grisms span from 0.8-2.5$\micron$ with resolution $R\sim200$.   
Most CALSPEC stars have counterpart theoretical spectra available as well \citep{2017AJ....153..234B}, and regions of the empirical spectra that lack observational coverage are filled in with model predictions where available.
We do not retain stars without empirical WFC3/IR and/or NICMOS data in our final analysis, but show them in \autoref{ssec:2mass_obs} for comparison.

In addition to CALSPEC's sub-percent absolute and relative calibration and widespread use as standards \citep[e.g. ][]{2006AJ....132.1221H, 2010A&A...512A..54C, 2012PASP..124..140B, 2012ApJ...750...99T, 2018ApJ...859..101S}, its lack of atmospheric absorption makes it an extraordinarily valuable anchor, as the other spectral libraries we use are all from ground-based observatories.

Throughout this work, we use the February 2023 CALSPEC release%\footnote{\url{https://web.archive.org/web/20230312225336/https://www.stsci.edu/hst/instrumentation/reference-data-for-calibration-and-tools/astronomical-catalogs/calspec}} %
, which includes the Vega spectrum \texttt{alpha\_lyr\_stis\_011.fits}.
We use this spectrum as our reference for the Vega magnitude system in all bands.

\subsubsection{IRTF and Extended IRTF Libraries} \label{sssec:irtf}

The original IRTF Spectral Library\footnote{\url{http://irtfweb.ifa.hawaii.edu/~spex/IRTF_Spectral_Library/}} \citep{2009ApJS..185..289R} contains 226 $R\sim2000$ spectra of 225 late-type stars at mostly solar metallicity, taken by the SpeX prism spectrograph at the NASA Infrared Telescope Facility on Mauna Kea \citep{2003PASP..115..362R} with coverage from 0.8-2.5$+$~$\micron$.
We adopt the data as presented in the library, which relied on the data processing tool described in \citet{2004PASP..116..362C}, the telluric correction described in \citet{2003PASP..115..389V}, and the uncertainty propagation described in \citet{2004PASP..116..352V}. 

The Extended IRTF Library (EIRTF, \citealt{2017ApJS..230...23V})\footnote{\url{http://irtfweb.ifa.hawaii.edu/~spex/IRTF_Extended_Spectral_Library/}} expands the metallicity range of the original library, adding spectra of 206 stars with $-1.7 <$ [Fe/H] $< 0.6$ taken with the same instrument.
Both phases of the IRTF libraries are flux calibrated to 2MASS photometry \citep[see][\S2.3]{2009ApJS..185..289R}.

\subsubsection{X-Shooter Spectral Library} \label{sssec:xsl}

The X-Shooter Spectral Library\footnote{\url{http://xsl.astro.unistra.fr/index.html}} \citep{2014AA...565A.117C, 2020AA...634A.133G, 2022AA...660A..34V} was designed to bridge the optical and NIR for studies of intermediate to old-aged stellar populations.
As such, it spans a wide range of spectral types and luminosity classes with an emphasis on cool giants, including first-ascent red giants, red supergiants, and both oxygen- and carbon-rich thermally-pulsing asymptotic giant branch (TP-AGB) stars \citep{2016A&A...589A..36G, 2018arXiv181102841L}.
As of the third data release \citep{2022AA...660A..34V}, the library comprises 830 spectra of 683 unique stars observed with the ultraviolet, visible, and near-IR arms of the VLT/X-shooter spectrograph simultaneously.
The final set of arm-merged spectra offer $R\sim10,000$ spectral coverage from 0.35 to 2.48~$\micron$.
Of these, 736 (89\%; 614 unique stars) are fully corrected for wavelength-dependent instrumental flux losses \citep[as described in][]{2020AA...634A.133G}.

Detailed comparisons between XSL and other empirical spectral libraries, as well as between synthetic and observed {\em{Gaia}} and 2MASS broadband colors, find overall agreement of 5\% or better for common non-variable stars in all cases, and 1\% or better in many \citep{2020AA...634A.133G, 2022AA...660A..34V}.
Comparisons with model atmospheres and literature fundamental parameters, however, highlight ongoing difficulties in reproducing both overall SED shapes and specific spectral features for stars at low temperatures ($T_{\rm{eff}} \lesssim 4000$~K) and surface gravities ($\log{g} \lesssim 1$) \citep{2017A&A...601A.141G, 2019A&A...627A.138A, 2021A&A...649A..97L}.
% XSL SSP: 2022AA...661A..50V

%%%%%%%%%%%%%%%%%%%%%%%%%%%%%%%%%%%%%%%%%%%%%%%%%%%%%%%%%%%%%%%%%%%%%%%%%%%%%%
\subsection{Distances and Reddenings} \label{ssec:distance_reddening}
%%%%%%%%%%%%%%%%%%%%%%%%%%%%%%%%%%%%%%%%%%%%%%%%%%%%%%%%%%%%%%%%%%%%%%%%%%%%%%

\begin{figure*}[htbp]
    \centering
    \includegraphics[width=0.9\textwidth]{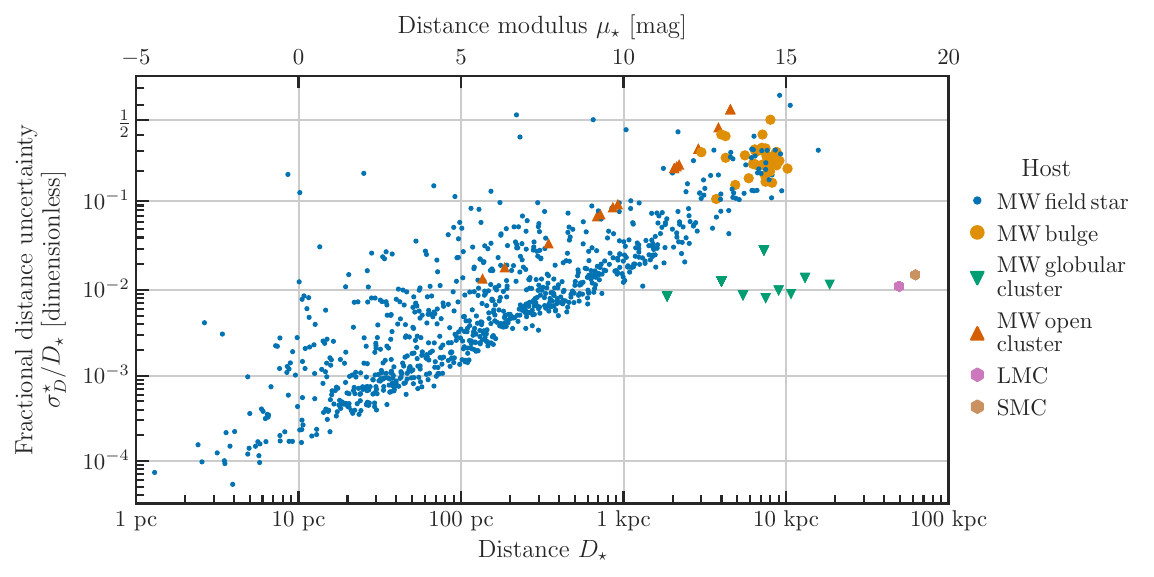}
    \caption{Fractional distance uncertainties as a function of distance (linear on the lower axis, modulus on the upper) for all library stars.
    Colors and symbols correspond to host objects, which are Galactic field or bulge stars, open and globular clusters, and the Magellanic Clouds.
    Note that many of the distance uncertainties are asymmetric; we show the mean values here.
    }
    \label{fig:distance_uncertainties}
\end{figure*}

While absolute-scale synthetic fluxes are neither necessary nor practical for our purposes, line-of-sight reddenings are important for interpreting synthetic colors.
All of the spectral libraries in \autoref{ssec:libraries} provide dereddened spectra with the exception of CALSPEC, which offers reddened model spectra instead (and whose targets are largely low-extinction regardless).
However, we opt to deredden all of the spectra as uniformly as possible using independent $E(B-V)$ estimates for inter-library consistency.

For the majority of Galactic stars, we are able to utilize three-dimensional dust maps to obtain the best estimate of reddening along the line-of-sight to the star.
Thus, the process of adopting reddening has two steps: first, the determination of the distance to the star and second, the estimation of the reddening from maps.

%%%%%%%%%%%%%%%%%%%%%%%%%%%%%%%%%%%%%%%%%%%%%%%%%%%%%%%%%%%%%%%%%%%%%%%%%%%%%%
\subsubsection{Distances} \label{sssec:distances}

For the vast majority of Galactic field stars we use geometric distances from \citet{2021AJ....161..147B}.
These distances use a Bayesian framework to invert the {\em{Gaia}} EDR3 trigonometric parallaxes and derive realistic uncertainties using a Galactic disk prior.
We find available distances for 1030 stars in this catalog, but 136 of these are superseded by distances to parent objects as described below, such that onlu 894 stars have distances adopted directly from \citet{2021AJ....161..147B}.

For stars that are members of Galactic stellar clusters, we use the mean distances derived in \citet{2020A&A...633A..99C} and \citet{2021MNRAS.505.5957B} for open and globular clusters respectively.
Both of these works combine astrometric and kinematic information for individual stars to determine aggregate cluster distances. 
\citet{2020A&A...633A..99C} provided catalogs with membership probabilities at a star-by-star level, and we adopt the cluster distance for library stars that have cluster membership probabilities greater than 50\%, which 17 stars in our sample do.
For globular cluster candidate stars, we determine cluster membership by first querying SIMBAD for hierarchical parent membership probabilities.
For stars with available membership data, we supersede the \citet{2021AJ....161..147B} distance if the median cluster membership probability is greater than 50\%.
For 9 candidates without membership data, we compare individual \citet{2021AJ....161..147B} distances to those derived for the parent clusters, and accept the cluster distances for all but two stars whose individual distances differ from the cluster distance by more than a factor of two.

For Magellanic Cloud stars, we do not require distance information to obtain reddening values, as 2D extinction maps offer sufficient correction.
For completeness, we adopt the detached eclipsing binary distances from \citet[][$m-M=18.477 \pm 0.004 \pm 0.026$]{2019Natur.567..200P} and \citet[][$m-M=18.977 \pm 0.016 \pm 0.028$]{2020ApJ...904...13G} for 51 and 25 stars in the Large and Small Magellanic Clouds respectively.

For the remaining 34 stars without distances from any of these sources, 14 have parallax distances from {\em{Gaia}} DR2 \citep{2018AJ....156...58B}, and 13 from {\em{Hipparcos}} \citep{2007A&A...474..653V}.
Seven have distances derived from various other methods, primarily luminosity fitting \citep{2007AJ....134.2200S, 2010AJ....139.1808S, 2010A&ARv..18...67T, 2020MNRAS.493..468D, 2022A&A...657A.131M}.

We were unable to locate any literature distance information for only 1 star, UCAC2 18505584.
As this star is towards the Galactic bulge, we conservatively assume its distance is 8 $\pm$ 4~kpc.

%%%%%%

In summary, distances for 894 stars were adopted from \citet{2021AJ....161..147B}, 57 stars from cluster distances, 76 from Magellanic Clouds eclipsing binary distances, and 34 from other sources.

\autoref{fig:distance_uncertainties} shows average fractional distance uncertainties ($\sigma^{\star}_{D} / D_{\star}$) against distance moduli ($\mu_{\star}$) for our preferred distances for all library stars.
Many of these distances have asymmetric upper and lower uncertainties; we  describe how we incorporate asymmetric distance uncertainties into our star-by-star $E(B-V)$ estimates where possible in \autoref{sssec:reddenings}.
For Galactic field and bulge stars, uncertainty largely scales with distance, as expected.
The best-measured of these are below 1\% distance uncertainty out to $\sim$1~kpc, and 10\% at $\sim$10~kpc.
The uncertainties on open cluster distances scale similarly, as we adopt the most conservative reported uncertainties based on assumed parallax offsets of $\Delta \omega = \pm 0.1$ mas  \citep{2020A&A...633A..99C}.
The globular cluster distances of \citet{2021MNRAS.505.5957B}, on the other hand, incorporate information from multiple distance determination techniques wherever available, of which parallax inversion is only one.
Several of these distances reach 1\% uncertainty out to $\sim$10~kpc, and all are within 5\%.
Finally, the detached eclipsing binary distances to the Large and Small Magellanic Clouds have 1- to 2\% formal uncertainties.
These are almost certainly underestimates for individual stars given line-of-sight depth variations, especially in the SMC \citep[][and references therein]{2017MNRAS.472..808R, 2022MNRAS.512..563R}.
However, the larger distance uncertainties are not a concern for reddening corrections, as the Magellanic Clouds reddening map is distance-independent.

%%%%%%%%%%%%%%%%%%%%%%%%%%%%%%%%%%%%%%%%%%%%%%%%%%%%%%%%%%%%%%%%%%%%%%%%%%%%%%
\subsubsection{Reddenings} \label{sssec:reddenings}

\begin{figure*}[htbp]
    \centering
    \includegraphics[width=0.9\textwidth]{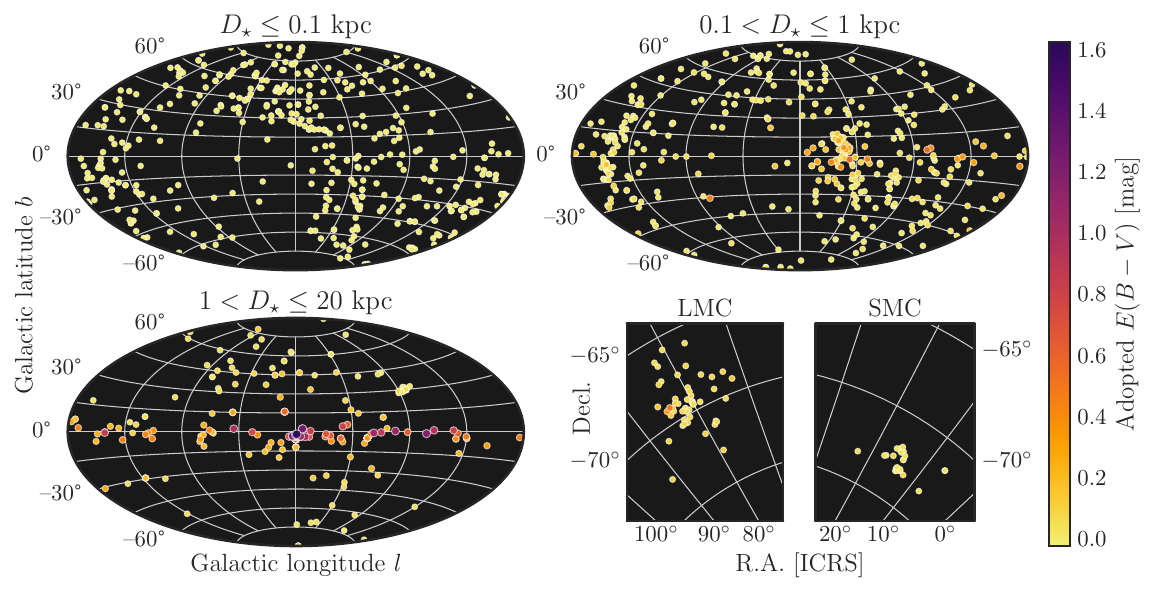}
    \caption{Sky maps of our final cross-library sample, color coded by adopted $E(B-V)$ values.
    We show all-sky maps in Galactic coordinates for Milky Way stars in three distance bins of 0-0.1~kpc, 0.1-1~kpc, and 1-20~kpc. 
    Magellanic Cloud stars are shown in separate insets in the lower right in equatorial coordinates.}
    \label{fig:ebv_allsky}
\end{figure*}

With distances in hand, we can now obtain line-of-sight reddening estimates for our sample.
For the majority of Galactic disk and bulge stars, including cluster stars, we use the 3D all-sky map from \citet{2022AA...664A.174V}\footnote{Available at \url{https://explore-platform.eu/sdas}.}, which uses a hierarchical inversion algorithm to derive line-of-sight extinctions over a $10\times10\times0.8$~kpc volume around the Sun to stars with high-quality \emph{Gaia} EDR3 parallaxes and \emph{Gaia} and 2MASS photometry.
This map revises \citet{2022AA...661A.147L} with an improved calibration for cool giant stars specifically, and \citet{2022AA...661A.147L} itself updates \citet{2019AA...625A.135L}.
We use this map for 808 stars in total.

For the 178 Galactic stars that fall outside the boundaries of the \citet{2022AA...664A.174V} map, we use the {\em{Gaia}} DR3 total Galactic extinction map \citep[TGE, ][]{2022arXiv220606710D}.
This map combines individual extinction measurements inferred from low-resolution BP/RP spectra \citep{2022arXiv220606138A} of giants outside the Galactic plane ($|b| > 300$~pc or Galactocentric radius $R > 16$~kpc).

Finally, for Magellanic Cloud stars we use the 2D map presented by \citet{2021ApJS..252...23S}, which is derived from OGLE-IV red clump color excess measurements.

In \autoref{app:extinction} we compare our final adopted $E(B-V)$ values both to those adopted or measured by the original libraries, and to those from several other widely used maps and stellar parameter fitting techniques as well.
We find that different $E(B-V)$ estimates have typical median absolute deviations (MADs) of 0.05-0.1~mag, with some exceptions.
This translates to a $\sim$2-5\% MAD in $E(J-K_S)$, assuming $E(J-K_S) / E(B-V) = 0.42$ \citep{2013MNRAS.430.2188Y}.

A similar intercomparison of the values from \citet{2021ApJS..252...23S} with other such measurements in the Magellanic Clouds \citep[e.g.][]{2007ApJ...662..969I, 2011AJ....141..158H, 2018ApJ...866...90C, 2019A&A...628A..51J, 2019MNRAS.489.3200B, 2022MNRAS.511.1317C} would be of potential interest as well.
We do not pursue it here because the proportion of Magellanic Clouds stars in our sample is low (below 10\%), and many of those spectra are affected by data quality issues regardless.

%%%%%%%%%%%%%%%%%%%%%%%%%%%%%%%%%%%%%%%%%%%%%%%%%%%%%%%%%%%%%%%%%%%%%%%%%%%%%%
\subsection{Global Sample Properties} \label{ssec:mixed_library}
%%%%%%%%%%%%%%%%%%%%%%%%%%%%%%%%%%%%%%%%%%%%%%%%%%%%%%%%%%%%%%%%%%%%%%%%%%%%%%

%%%%%%%%%%%%%%%%%%%%%%%%%%%%%%%%%%%%%%%%%%%%%%%%%%%%%%%%%%%%%%%%%%%%%
\begin{figure}
    \centering
    \includegraphics[width=\columnwidth]{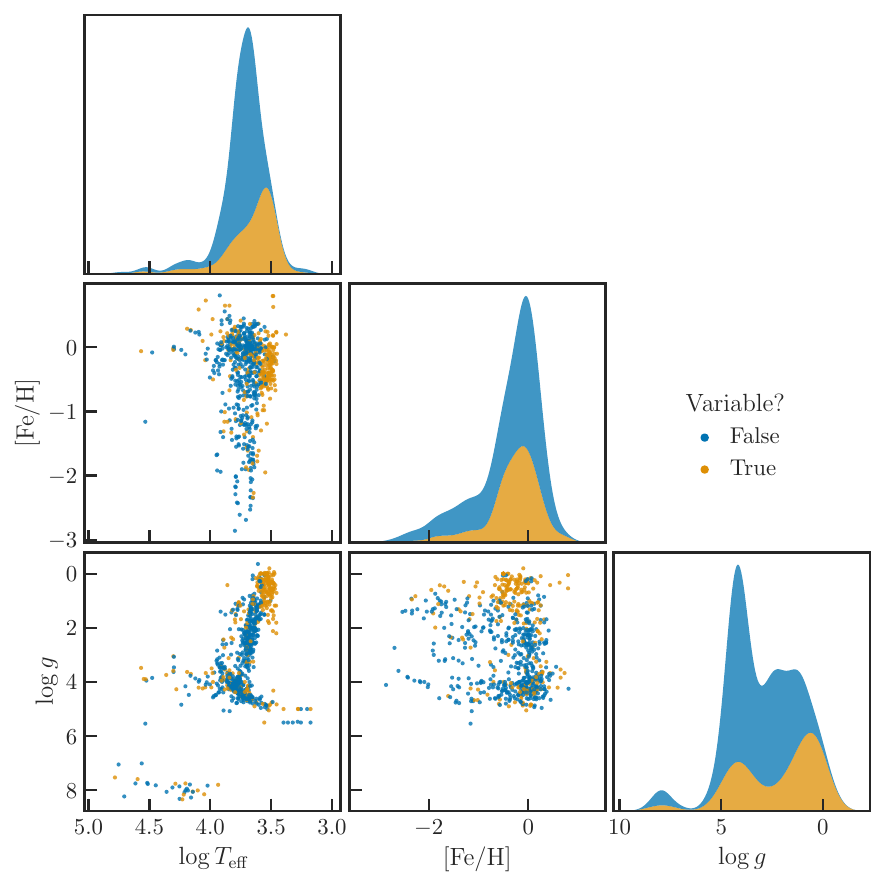}
    \caption{Summary stellar properties (effective temperature $T_{\rm eff}$, 
        metallicity [Fe/H], 
        and surface gravity $\log(g)$) of the merged spectral libraries. 
    We note that these literature measurements are extremely heterogenous, and these data are used primarily to understand the ranges of stellar properties over which our library can provide insight.
    Stars with known photometric variability are plotted in orange, whereas stars not known to be variable are shown in blue.}
    \label{fig:library_metadata}
\end{figure}
%%%%%%%%%%%%%%%%%%%%%%%%%%%%%%%%%%%%%%%%%%%%%%%%%%%%%%%%%%%%%%%%%%%%%

In addition to distances and reddenings, we have compiled relevant literature data for our sample from Vizier and SIMBAD.
These include stellar parameters such as effective temperature $\Teff$, metallicity [Fe/H], and surface gravity $\logg$; observed {\em{Gaia}} and 2MASS photometry; spectral type and other classifications; and binarity and variability information.
We show a summary of $\Teff$, $\logg$, and [Fe/H] coverage in \autoref{fig:library_metadata}.

\section{Synthetic Photometry} \label{sec:analysis}
%%%%%%%%%%%%%%%%%%%%%%%%%%%%%%%%%%%%%%%%%%%%%%%%%%%%%%%%%%%%%%%%%%%%%%%%%%%%%%
%%%%%%%%%%%%%%%%%%%%%%%%%%%%%%%%%%%%%%%%%%%%%%%%%%%%%%%%%%%%%%%%%%%%% 
Here we describe the process of producing and validating synthetic photometry from our ensemble library.
For ease of reproducing or expanding on this work, all derived products have been made public; we describe those data in \autoref{app:tables}. 

\subsection{Synthetic Photometry Procedure} \label{ssec:recipe}

For a given spectrum with flux density $f_\lambda(\lambda)$ and bandpass with dimensionless throughput $P(\lambda)$, the integrated flux density is defined as:
\begin{align} \label{eqn:flux_integration}
    f_\lambda(P) &\equiv \frac{\int f_\lambda(\lambda) P(\lambda) \lambda \, \mathrm{d}\lambda}{\int P(\lambda) \lambda \, \mathrm{d}\lambda}
\end{align}
\citep{hellerich1937handbuch, 1986HiA.....7..833K}.

If we treat the spectrum $f_\lambda(\lambda)$ as a vector $\mathbf{f_{\lambda}}$, and a set of bandpasses $\mathbf{P}$ as a matrix with dimensions $n_P \times n_{\lambda}$, where $n_P$ is the number of bandpasses and $n_{\lambda}$ is the number of wavelength samples, then
the integrated fluxes $\mathbf{f}_P$ 
are simply:
\begin{equation} \label{eqn:flux_dot}
    \mathbf{f}_P = \mathbf{P} \cdot \mathbf{f}_\lambda
\end{equation}

This approach allows us to calculate the covariance between integrated fluxes in different bandpasses as
\begin{equation} \label{eqn:flux_cov}
    \Cov[{\mathbf{f}_P}] = \mathbf{P} \cdot \Cov[{\mathbf{f}_\lambda}] \cdot \mathbf{P}^T
\end{equation}
following \citet{2022arXiv220606215G}.

We begin by resampling all spectra and transmission curves to the $R=100$ wavelength grid on which the \hst flux uncertainty scale is defined \citep{2014PASP..126..711B}, to improve computational efficiency and ease of uncertainty propagation.
We used the \texttt{FluxConservingResampler} from the Python package \texttt{specutils}, which preserves integrated flux and associated uncertainties \citep{2017arXiv170505165C}.
For XSL spectra, we first fill in the deepest telluric-dominated regions (1.35 - 1.41 and 1.81 - 1.935 $\mu$m) via convolution with a Gaussian kernel.

Next, we correct the resampled spectra for interstellar extinction.
Specifically, we estimate the total wavelength-dependent line-of-sight extinction $A_\lambda$ using the $E(B-V)$ values described in \autoref{sssec:reddenings} as inputs to extinction laws implemented in the \texttt{dust-extinction} package.\footnote{\url{https://web.archive.org/web/20230706195114/https://learn.astropy.org/tutorials/color-excess.html}}
We use the \citet{2019ApJ...886..108F} law with $R_V = 3.1$ for Galactic stars, and in the LMC and SMC we use the respective average laws from \citet{2003ApJ...594..279G}.
We incorporate both the reported uncertainties on $E(B-V)$ and, in the Galactic case, uncertainties of $\pm 0.3$ on $R_V$ into the final uncertainties on the deextinguished spectra.

Using equations~\ref{eqn:flux_dot} and \ref{eqn:flux_cov}, we are able to derive integrated fluxes and covariances for an arbitrary set of bandpasses $\mathbf{P}$ over both the original and deextinguished spectra.
For the remainder of this paper, we will focus primarily on the 2MASS $JHK_S$ and \HST WFC3/IR F110W, F125W, and F160W bandpasses.
However, we also provide synthetic {\em{Gaia}} DR3 $BP$, $G$, and $RP$ magnitudes for spectra with sufficient wavelength coverage in this paper. 
Comparable sets of transformations for {\em{JWST}}/NIRCam, {\em{Roman}}/WFI, and other ground-based systems will be presented in a follow-up paper.

\edit1{In this work we use throughput data provided under the \HST calibration reference data system (CRDS)\footnote{\url{https://hst-crds.stsci.edu/}} context \texttt{hst\_synphot\_0055.imap}.
For \HST these include a number of telescope and instrument component files which may be assembled into full-system throughputs using the \texttt{stsynphot} Python package \citep{2020ascl.soft10003S}.
The WFC3 implementation offers options to specify an observation date and aperture radius, which modify the net throughputs using encircled energy and time-dependent sensitivity information \citep{2009wfc..rept...37H, 2022AJ....164...32C}; here we use the default ``infinite" (6.0\arcsec) aperture and instrument reference epoch MJD=55008.
The canonical 2MASS relative spectral response curves \citep{2003AJ....126.1090C} are also provided in CRDS for use with \texttt{stsynphot}.
The relevant transmission curve files are named \texttt{wfc3\_ir\_{$<$}band{$>$}\_mjd\_007\_syn.fits} and \texttt{2mass\_{$<$}band{$>$}\_001\_syn.fits} respectively.
}

%%%%%%%%%%%%%%%%%%%%%%%%%%%%%%%%%%%%%%%%%%%%%%%%%%%%%%%%%%%%%%%%%%%%%
\subsection{Validation} \label{ssec:2mass_obs}
%%%%%%%%%%%%%%%%%%%%%%%%%%%%%%%%%%%%%%%%%%%%%%%%%%%%%%%%%%%%%%%%%%%%%

Overall, the best check on synthetic magnitudes and colors is to compare them to corresponding measured values.
In this section, we first cross-check our CALSPEC data against high-quality 2MASS and {\em{Gaia}} photometry, and then consider the global 2MASS color-magnitude and color-color fidelity of the full sample. 

\subsubsection{CALSPEC zeropoints} \label{sssec:zp}

%%%%%%%%%%%%%%%%%%%%%%%%%%%%%%%%%%%%%%%%%%%%%%%%%%%%%%%%%%%%%%%%%%%%%
\begin{figure*}
    \centering
    \includegraphics[width=\textwidth]{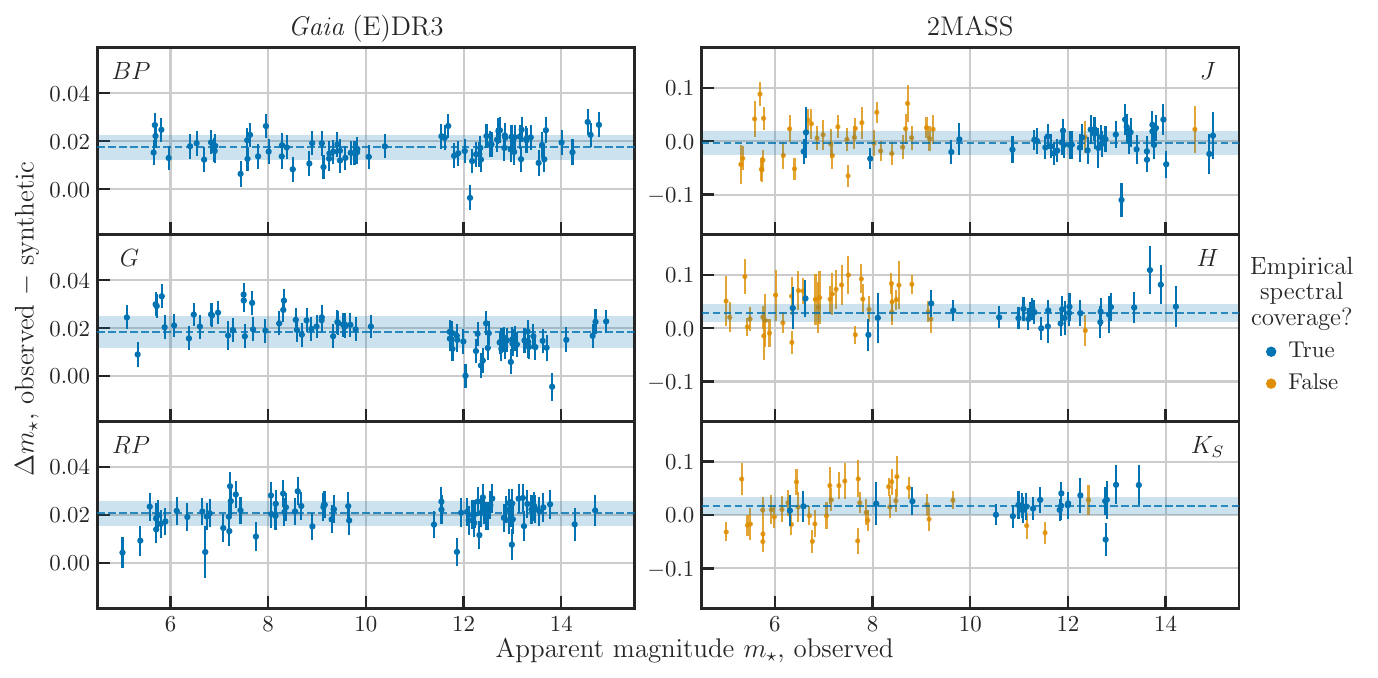}
    \caption{Differences between observed and synthetic magnitudes in {\em{Gaia}} (E)DR3 $BP$, $G$, $RP$ (left) and 2MASS $JHK_S$ (right) as functions of observed magnitude for select CALSPEC stars.
    Blue points indicate empirical spectrophotometry (available for effectively all CALSPEC stars in the optical), and orange indicates spectra extrapolated from theoretical models (e.g. missing WFC3/IR and/or NICMOS data), which we exclude from calculations and show only for reference. 
    Blue dashed lines and solid bands show weighted means and standard deviations of the photometric offsets, as reported in \autoref{tab:calspec}.
    }
    \label{fig:calspec_comparison}
\end{figure*}
%%%%%%%%%%%%%%%%%%%%%%%%%%%%%%%%%%%%%%%%%%%%%%%%%%%%%%%%%%%%%%%%%%%%%

\defcitealias{2018A&A...616L...7M}{MA18}

While an end-to-end recalibration in either direction \citep[as in, e.g., ][]{2011wfc..rept...15R, 2018A&A...619A.180M, 2018A&A...616L...7M, 2022arXiv220606215G} is beyond the scope of this paper, here we briefly reproduce the 2MASS zeropoint adjustment procedure of \citet[][hereafter MA18]{2018A&A...616L...7M} and references therein.
We extend this procedure to {\em{Gaia}} (E)DR3 as well for additional verification, as in \citet[][section~4.4.2]{2021A&A...649A...5F}.

We choose our comparison sample based on the criteria of \citetalias{2018A&A...616L...7M} with minor updates, namely:
a) CALSPEC spectra with empirical IR spectrophotometry (limited to NICMOS in \citetalias{2018A&A...616L...7M}; here we also include WFC3/IR); and
b) uncertainties on observed magnitudes of 0.05 mag or less.
For {\em{Gaia}}, we use all available CALSPEC stars with observed magnitudes $5 < m_\star < 15$.
All criteria are applied on a per-band basis.

We calculate mean offsets and dispersions weighted by combined synthetic and observed magnitude uncertainties for each band, as summarized in \autoref{tab:calspec} and shown in \autoref{fig:calspec_comparison}.
We find a consistent $\sim$0.02~mag offset between the synthetic and observed photometry in all {\em{Gaia}} bands, and offsets of up to 0.03~mag in 2MASS, although the latter are all statistically compatible with zero.
We attribute these largely to differences in the respective adopted reference SEDs, consistent with previous studies of synthetic {\em{Gaia}} and 2MASS magnitudes (\citetalias{2018A&A...616L...7M}, \citealt{2018MNRAS.479L.102C}) and historical changes to the CALSPEC Vega calibration \citep{2014PASP..126..711B, 2019AJ....158..211B}.

\begin{table}[ht]
\centering
\begin{tabular}{c@{}c|rcccc}
\hline
\multicolumn{2}{c|}{Band} &  {$\Delta$ZP} &  {$\sigma$($\Delta$ZP)}& {$\bar{\sigma}_\mathrm{obs}$} & {$\bar{\sigma}_\mathrm{syn}$} & {$N_\star$} \\ \hline \hline
{\em{Gaia}} &  $BP$ &  0.0175 & 0.0052 & 0.0030 & 0.0042 & 80 \\
            &   $G$ &  0.0184 & 0.0068 & 0.0028 & 0.0042 & 82 \\
            &  $RP$ &  0.0208 & 0.0052 & 0.0039 & 0.0045 & 80 \\ \hline
   {2MASS} &   $J$ & --0.0030 & 0.0229 & 0.0248 & 0.0085 & 45 \\
            &   $H$ &  0.0283 & 0.0172 & 0.0258 & 0.0099 & 30 \\
            & $K_S$ &  0.0168 & 0.0176 & 0.0244 & 0.0116 & 23 \\ \hline \hline
\end{tabular}
\caption{Summary of Gaia (E)DR3 and 2MASS observed and synthetic photometry differences for CALSPEC stars, following the method of \citetalias{2018A&A...616L...7M} and references therein.
$\Delta$ZP and $\sigma$($\Delta$ZP) are the error-weighted mean and standard deviation of the differences (in the direction observed -- synthetic); $\bar{\sigma}_\mathrm{obs}$ and $\bar{\sigma}_\mathrm{syn}$ are the respective mean uncertainties; and $N_\star$ is the number of stars used.
}
\label{tab:calspec}
\end{table}

\subsubsection{Global CMD and color-color fidelity} \label{sssec:cmds}
%%%%%%%%%%%%%%%%%%%%%%%%%%%%%%%%%%%%%%%%%%%%%%%%%%%%%%%%%%%%%%%%%%%%%
\begin{figure*}
    \centering
    \includegraphics[width=0.95\textwidth]{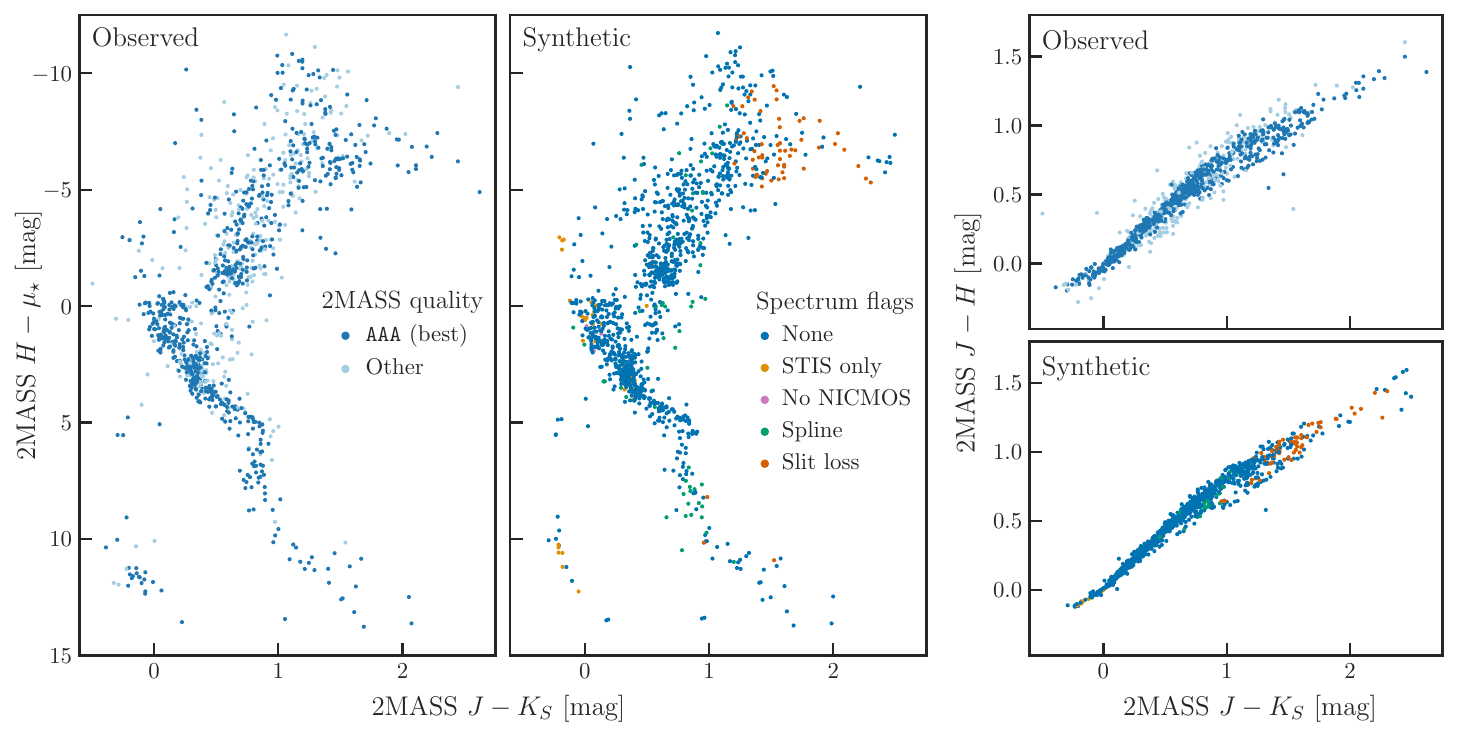}
    \caption{
    Left: absolute $H$ vs.\ $J-K_S$ CMD of all available 2MASS photometry for our full sample. 
    Here, dark blue points indicate sources with high-quality 2MASS photometry in all bands (``\texttt{Qflg=AAA}"), and light blue indicates non-optimal quality flags in one or more bands. 
    Center: the same CMD for our synthetic photometry suite.
    Blue points indicate spectra with no data caveats.
    Orange indicates CALSPEC library stars that were observed only with STIS, where the spectra are filled in outside the STIS wavelength range ($\gtrsim 1~\micron$) using theoretical spectra.
    Green indicates XSL spectra that were corrected for slit loss via spline fitting \citep{2022AA...660A..34V}, and red indicates XSL spectra uncorrected for slit loss.
    Right: The same in $J-H$ vs.\ $J-K_S$ color-color space.
    }
    \label{fig:obs_syn_abscmds}
\end{figure*}
%%%%%%%%%%%%%%%%%%%%%%%%%%%%%%%%%%%%%%%%%%%%%%%%%%%%%%%%%%%%%%%%%%%%%

\autoref{fig:obs_syn_abscmds} compares the absolute observed 2MASS magnitudes (left) to their synthetic counterparts (right) in a $M_{H}$ versus $J-H$ color magnitude diagram.
The right panel of \autoref{fig:obs_syn_abscmds} separates the highest quality photometry, \texttt{Qflg~=~AAA} \citep{2006AJ....131.1163S}, from that with flags in one or more bands; the best photometry is shown as dark blue, and the lower quality in lighter blue. 
Visually, the stars with lower photometric quality have larger scatter around prominent CMD features (e.g., the red giant branch, main sequence, and white dwarf cooling sequence) that are tightly traced by the high-quality photometry.

The left panel of \autoref{fig:obs_syn_abscmds} uses our synthetic magnitudes, where the points are color-coded if there are quality flags within the spectral libraries themselves.
Of note: 
    the orange points are CALSPEC stars observed only with STIS, meaning that the spectrum beyond $\sim\!1~\micron$ is synthetic \citep{2017AJ....153..234B};
    the green are XSL spectra corrected for slit-dependent flux loss with a spline function \citep[][]{2022AA...660A..34V};
    and the red are XSL spectra uncorrected for slit loss. 

Generally, the scatter in the right panel of \autoref{fig:obs_syn_abscmds} is reduced with the synthetic photometry, which is a comforting check on our processes.
However, we emphasize that the absolute-scale synthetic magnitudes for XSL in particular should be treated with caution, as the XSL spectra are calibrated only for relative (wavelength-dependent) flux and not absolutely fluxed.
We do not make more detailed comparisons here because of the relatively large sample of stars known to be variable, particularly on the upper giant branch (e.g., \autoref{fig:library_metadata}) and the comparison here would likely only serve to highlight cases where the spectra and photometry were taken at inconsistent phase points.

\section{Results} \label{sec:res}
%%%%%%%%%%%%%%%%%%%%%%%%%%%%%%%%%%%%%%%%%%%%%%%%%%%%%%%%%%%%%%%%%%%%%%%%%%%%%%
\subsection{Fitting sample selection} \label{ssec:fitting_sample}

\begin{figure}
    \centering
    \includegraphics[width=\columnwidth]{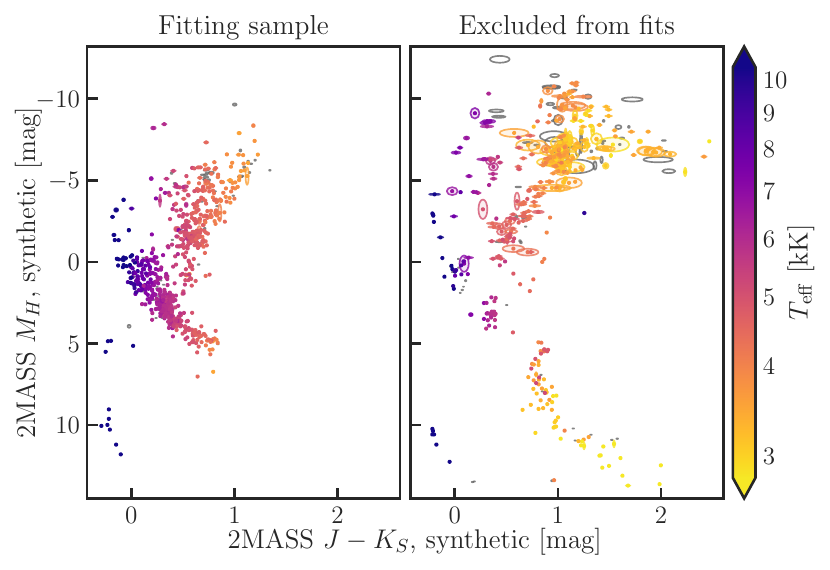}
    \caption{Color-magnitude diagrams of our final fitting sample (left) and stars excluded from fitting (right), colormapped by literature effective temperature.}
    \label{fig:fitting_selection}
\end{figure}

To construct general-purpose filter transformations, we require a subset of library stars with high-quality spectra covering as broad a swath of the HR diagram as possible.
Given the heterogeneity of available information for stars in our library, we combine several complementary kinds of classification data to accept or reject stars for our final calibration sample. 
We selected the fitting sample based on the following criteria:
\begin{enumerate}
    \item Synthetic magnitude uncertainties under 0.05 mag in all bands;
    \item No spectral quality flags (see \autoref{fig:obs_syn_abscmds});
    \item No known atypical SED features, based on available SIMBAD object class and spectral type information.
    Specifically, we reject stars with any of the following literature classifications: 
    \begin{enumerate}
        \item Ae or Be stars, which often show IR excesses from circumstellar decretion disks;
        \item dwarves of type M or later, including L and T dwarves;
        \item C-, R-, and S-type stars; these are mainly AGB stars of varying masses, but also include carbon-enhanced metal poor (CEMP) and other chemically peculiar non-AGB stars;
        \item other extreme giant types, including Miras, OH/IR and post-AGB stars, and supergiants of any temperature;
        \item any other stars with $T_\mathrm{eff} < 3500$~K.
    \end{enumerate}
\end{enumerate}

Together, these yield a sample of 773 high-quality spectra of 648 unique stars in total.
Color-magnitude diagrams of accepted and rejected stars are shown in \autoref{fig:fitting_selection}.
The fitting sample is dominated by main sequence and red giant branch stars the color range $-0.3 < J-K_S < 1.35$.

\subsection{Fitting process} \label{ssec:fitting}

\begin{table*}
    \centering
\caption{Filter transformation results} \label{tab:fitting}
    \begin{tabular}{c c R R R R c c c}
    \hline \hline
$Y$ & $X$ & $X_0 & X_1$ & \multicolumn{1}{C}{c_0} & \multicolumn{1}{C}{c_1} & $\sigma_\mathrm{seg}$ & $N_\star$ & $\sigma_\mathrm{fit}$ \\ \hline
F110W -- $J$ &          $J-H$ & -0.13 &  0.20 & -0.0066 &  0.4743 &  0.0091 &  153 &  0.0088 \\
             &                &  0.20 &  1.02 &  0.0396 &  0.2419 &  0.0087 &  620 &      '' \\
             &        $J-K_S$ & -0.29 & -0.04 & -0.0103 &  0.2244 &  0.0078 &   29 &  0.0088 \\
             &                & -0.04 &  0.24 & -0.0025 &  0.4065 &  0.0117 &  137 &      '' \\
             &                &  0.24 &  1.34 &  0.0494 &  0.1885 &  0.0080 &  607 &      '' \\
             & F110W -- F160W & -0.17 &  0.21 & -0.0069 &  0.3797 &  0.0059 &  125 &  0.0073 \\
             &                &  0.21 &  1.09 &  0.0200 &  0.2518 &  0.0075 &  648 &      '' \\
             & F125W -- F160W & -0.09 &  0.14 & -0.0096 &  0.6162 &  0.0090 &  128 &  0.0093 \\
             &                &  0.14 &  0.76 &  0.0278 &  0.3404 &  0.0094 &  645 &      '' \\
\hline
F125W -- $J$ &          $J-H$ & -0.13 &  1.02 & -0.0023 & -0.0115 &  0.0050 &  773 &  0.0050 \\
             &        $J-K_S$ & -0.29 &  1.34 & -0.0027 & -0.0090 &  0.0051 &  773 &  0.0051 \\
             & F110W -- F160W & -0.17 &  1.09 & -0.0018 & -0.0111 &  0.0051 &  773 &  0.0051 \\
             & F125W -- F160W & -0.09 &  0.76 & -0.0018 & -0.0156 &  0.0051 &  773 &  0.0051 \\
\hline
F160W -- $H$ &          $J-H$ & -0.13 &  0.31 & -0.0080 &  0.1845 &  0.0063 &  268 &  0.0081 \\
             &                &  0.31 &  0.63 & -0.0330 &  0.2659 &  0.0089 &  386 &      '' \\
             &                &  0.63 &  1.02 & -0.0704 &  0.3256 &  0.0094 &  119 &      '' \\
             &        $J-K_S$ & -0.29 &  0.30 & -0.0027 &  0.1402 &  0.0064 &  217 &  0.0082 \\
             &                &  0.30 &  1.34 & -0.0257 &  0.2166 &  0.0089 &  556 &      '' \\
             & F110W -- F160W & -0.17 &  0.34 & -0.0079 &  0.1442 &  0.0076 &  224 &  0.0096 \\
             &                &  0.34 &  0.66 & -0.0496 &  0.2668 &  0.0100 &  407 &      '' \\
             &                &  0.66 &  1.09 & -0.0992 &  0.3424 &  0.0113 &  142 &      '' \\
             & F125W -- F160W & -0.09 &  0.24 & -0.0087 &  0.2253 &  0.0074 &  246 &  0.0102 \\
             &                &  0.24 &  0.49 & -0.0420 &  0.3634 &  0.0108 &  409 &      '' \\
             &                &  0.49 &  0.76 & -0.1022 &  0.4864 &  0.0131 &  118 &      '' \\  
\hline \hline
    \end{tabular}
\tablecomments{Filter transformations take the functional form $Y = c_0 + c_1 X$ for $X_0 < X \le X_1$, where $X$ is a broadband color and $Y$ is the difference between comparable filters.
$\sigma_\mathrm{seg}$ is the weighted RMS dispersion of the residuals in each segment (e.g. with $X_0 < X \le X_1$), and $\sigma_\mathrm{fit}$ is the same for the full piecewise relation. $N_\star$ is the number of stars per segment.} 
\end{table*}

\begin{figure*}
    \centering
    \includegraphics[width=\textwidth]{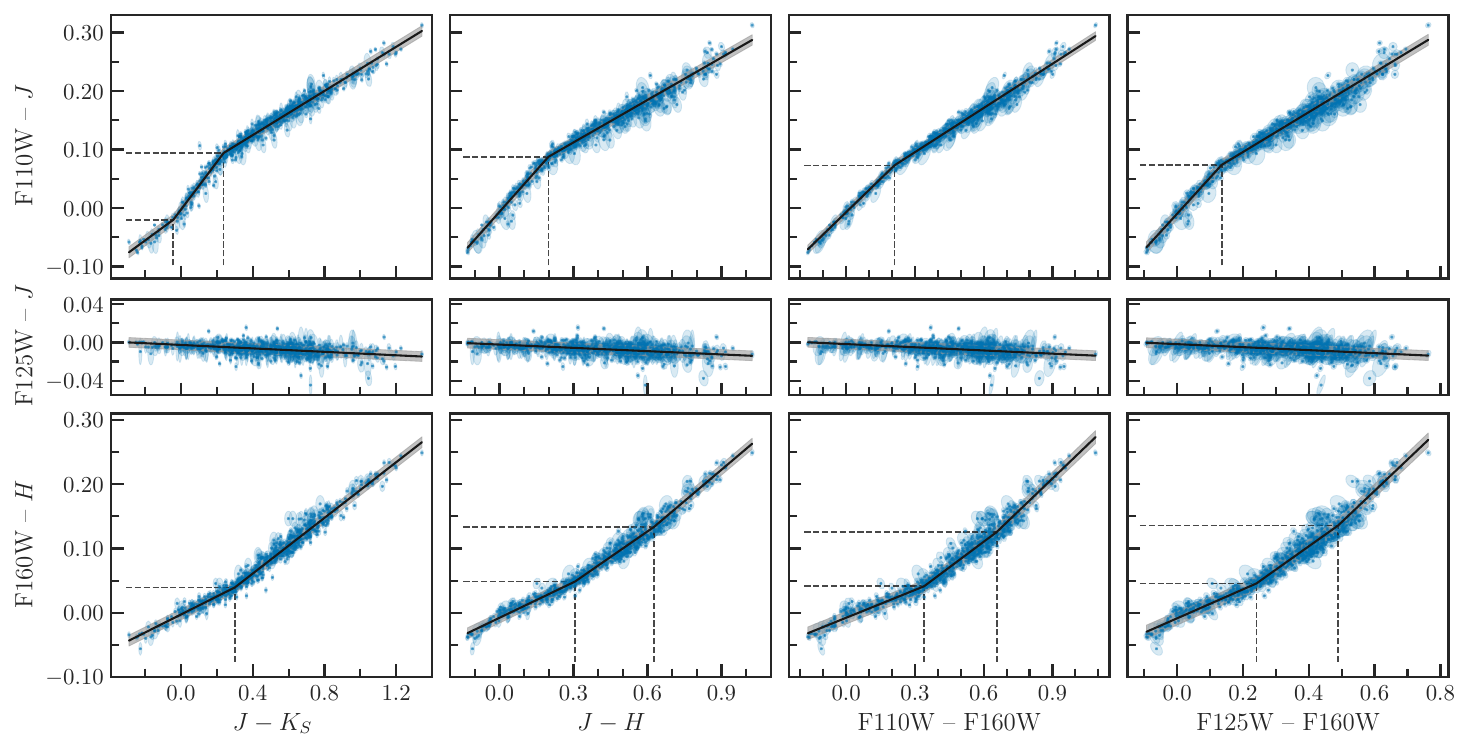}
    \caption{Filter transformation fits to F110W -- $J$ (top row), F125W -- $J$ (center row), and F160W -- $H$ (bottom row) as functions of $J-K_S$, $J-H$, F110W--F160W, and F125W--F160W color from left to right.
    Points used in the fitting are shown in dark blue with light blue covariance ellipses.
    Solid black lines show the resulting piecewise linear transformations with 68\% confidence intervals in grey, and dashed black lines mark the breakpoint locations.
    }
    \label{fig:fitting}
\end{figure*}

We plot differences between comparable \hst and 2MASS magnitudes as functions of color for our fitting sample in \autoref{fig:fitting}.
In general, they follow consistent and well-defined relations, albeit not necessarily linear ones.

Deviations from linearity are most clearly seen in F110W -- $J$ vs.\@ $J-K_S$ (the upper leftmost panel of \autoref{fig:fitting}).
Here, the noticeably steeper slope at $0 \lesssim J-K_S \lesssim 0.25$ corresponds to the $0.82~\micron$ Paschen jump, which is strongest in A-type stars \citep{1986ApJS...60..577T, 1998BaltA...7..571S}.
The slope change at the blue end of this feature is only detectable in F110W -- $J$ vs.\@ $J-K_S$, as F110W is the most blue-sensitive filter and $J-K_S$ the wides available color baseline.
However, we consistently observe slope changes at colors of about 0.2 to 0.35 mag in F110W -- $J$ and F160W -- $H$ for all color baselines.

A more subtle change in F160W -- $H$ also appears near 0.5 - 0.65 mag for all colors except $J-K_S$.
We attribute this to the $1.6~\micron$ ``bump", where the $H-$ continuum opacity reaches a minimum as the dominant absorption mode transitions from photoionization to free-free \citep{1988A&A...193..189J, 2023Atoms..11...61A}.
This produces a well-defined peak in cool stars' SEDs, which is strongest in intermediate M-type giants and supergiants \citep{2009ApJS..185..289R}. 
The $H$ band is nearly centered on this feature, but F160W falls about $0.1~\micron$ bluewards of its peak, resulting in slightly dimmer F160W magnitudes relative to $H$. 

We fit transformations using the Python library \texttt{pwlf} \citep{pwlf}, which solves for a system of piecewise linear equations continuous over the data domain.\footnote{For a detailed description of the \texttt{pwlf} algorithm and implementation, please see \url{https://web.archive.org/web/20211203035946/https://jekel.me/2018/Continous-piecewise-linear-regression/}.
The original description of the method may be found at \url{https://web.archive.org/web/20171116164225/https://www.golovchenko.org/docs/ContinuousPiecewiseLinearFit.pdf}}
The current version as of this writing, v0.2.0, offers the options to fit for optimal breakpoint locations given a desired number of line segments, and to incorporate weights in the least-squares regression.
For a synthetic color $X$ and \HST-2MASS magnitude difference $Y$, we weight by the inverse variance 
\begin{equation}
        w_{X,Y}^{-2} = \sigma_X^2 + \sigma_Y^2 - \mathrm{cov}({X,Y}) \label{eq:weights}\\
\end{equation}
where $\mathrm{cov}({X,Y})$ is obtained by propagating the uncertainties from \autoref{eqn:flux_cov}.

Our fitting results are summarized in \autoref{tab:fitting}, and overplotted in black in \autoref{fig:fitting}.

%%%%%%%%%%%%%%%%%%%%%%%%%%%%%%%%%%%%%%%%%%%%%%%%%%%%%%%%%%%%%%%%%%%%%%%%%%%%%%
\section{Discussion} \label{sec:dis}
%%%%%%%%%%%%%%%%%%%%%%%%%%%%%%%%%%%%%%%%%%%%%%%%%%%%%%%%%%%%%%%%%%%%%%%%%%%%%%
\subsection{Comparison with Literature Relations} \label{ssec:literature}

Earlier works have carried out similar exercises to what we have done here using theoretical stellar atmospheres. 

\citet{2011wfc..rept...15R} derived first-order transformations between F125W, F160W and $JH$ as part of an investigation of the WFC3/IR count rate nonlinearity (CRNL, also known as reciprocity failure).
They used synthetic photometry of \citet{2003IAUS..210P.A20C} model atmospheres with $-0.15 < J-H < 0.8$, $T_\mathrm{eff} \geq 3500$~K, $\logg$=4.5, and solar metallicity and abundance ratios.
With these they fit the following color terms:
\begin{align} 
        J - \F{125W} &= +0.012(\pm0.020) (J-H) \label{eq:R11_syn_j} \\
        H - \F{160W} &= -0.204(\pm0.040) (J-H) \label{eq:R11_syn_h} %\\
\end{align}

To interpret tip of the red giant branch (TRGB) color-magnitude absolute calibrations, \citet{2012ApJS..198....6D} derived transformations using Padova isochrone predictions \citep{2008PASP..120..583G} for 10~Gyr stars at the TRGB with $0.6 < J-H < 1.15$. 
Within this color range, the model transformations between $JH$ and F110W, F160W were fit by:
\begin{align}
    \begin{split}
        \F{160W} - H &= 0.2031 + 0.401(J-H-0.9) \label{eq:D12_h}\\
                     &\quad + 0.3498(J-H-0.9)^2 \\
    \end{split} \\
    \begin{split}
        J - H &= 0.9418 + 0.841(\F{110W}-\F{160W}-1.0) \label{eq:D12_color}\\
                     &\quad - 0.9053(\F{110W}-\F{160W}-1.0)^2 \\
    \end{split}
\end{align}

\begin{figure}
    \centering
    \includegraphics[width=\columnwidth]{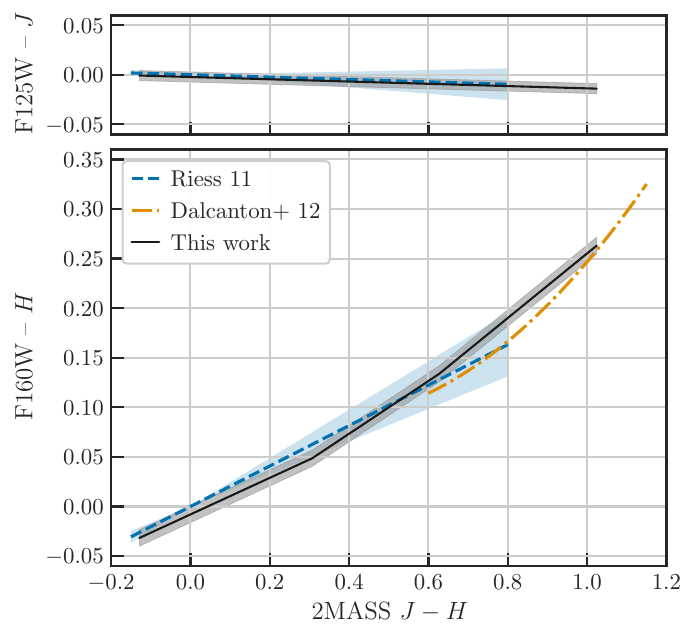}
    \caption{Comparisons to previously published transformations for F125W~--~$J$ \citep[][upper panel]{2011wfc..rept...15R} and F160W~--~$H$ (lower panel; \citealt{2011wfc..rept...15R} in blue, and \citealt{2012ApJS..198....6D} in orange).
    Results of this work are shown in gray.}
    \label{fig:lit_comparison}
\end{figure}

We show comparisons between equations~\ref{eq:R11_syn_j}-\ref{eq:D12_h} and our results in \autoref{fig:lit_comparison}.
We find overall excellent agreement with \citet{2011wfc..rept...15R} in both bands, and only a very slight offset relative to \citet{2012ApJS..198....6D} in F160W -- $H$ ($\sim$0.02~mag at $J-H = 0.8$, but less at both bluer and redder colors).

\subsection{Stellar properties} \label{ssec:residuals}

\begin{figure*}
    \centering
    \includegraphics[width=0.9\textwidth]{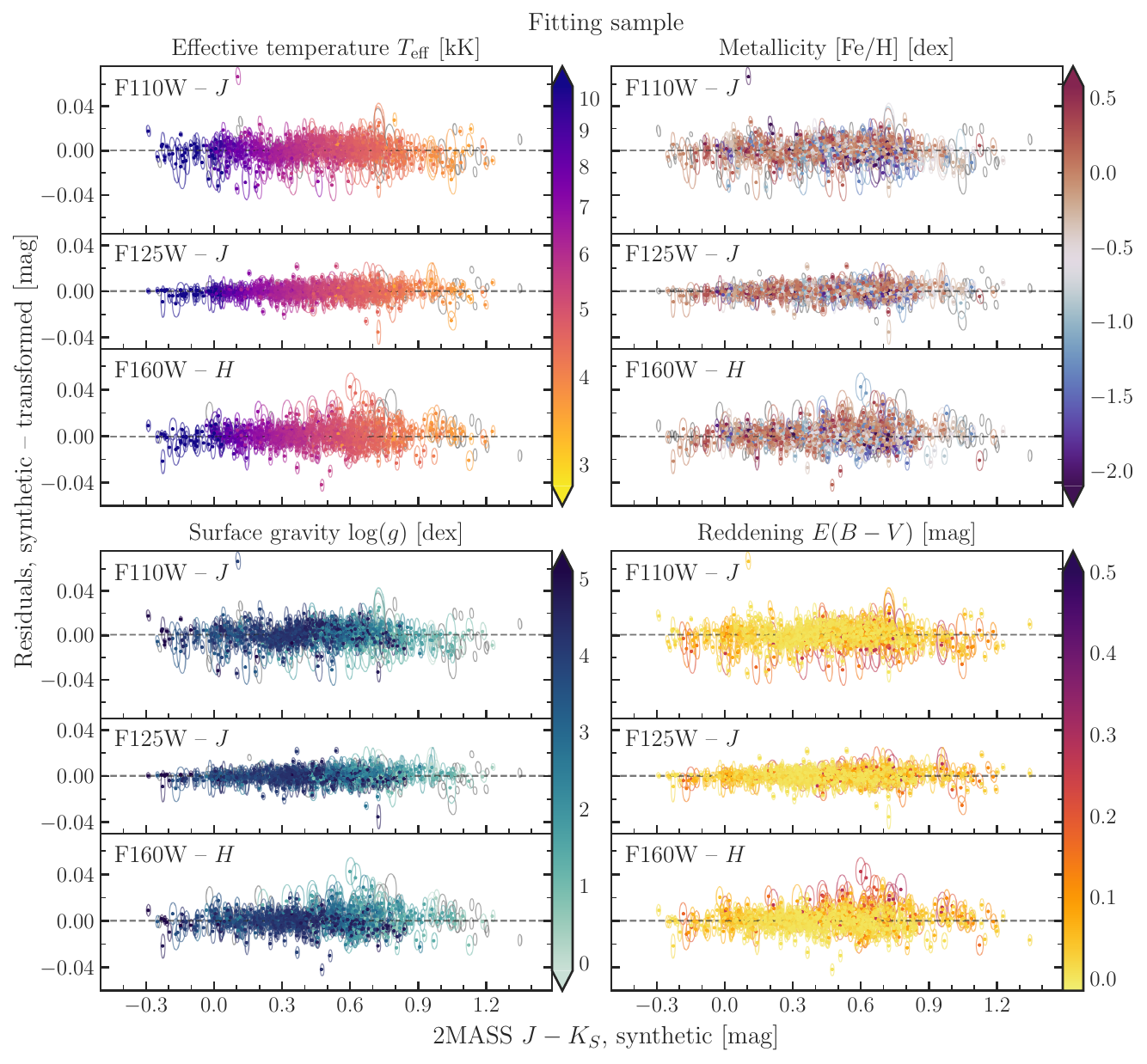}
    \caption{Residuals of our \HST-2MASS filter transformations with respect to $J-K_S$ color for our fitting sample (773 spectra of 648 unique stars).
    Colormaps show relevant stellar parameters where available, including effective temperature $\teff$, metallicity [Fe/H], surface gravity log($g$), and reddening $E(B-V)$.
    Reddenings are as described in \autoref{sssec:reddenings}; all other parameters are adopted as-is from literature sources.}
    \label{fig:jk_resid_fit}
\end{figure*}

\begin{figure*}
    \centering
    \includegraphics[width=0.9\textwidth]{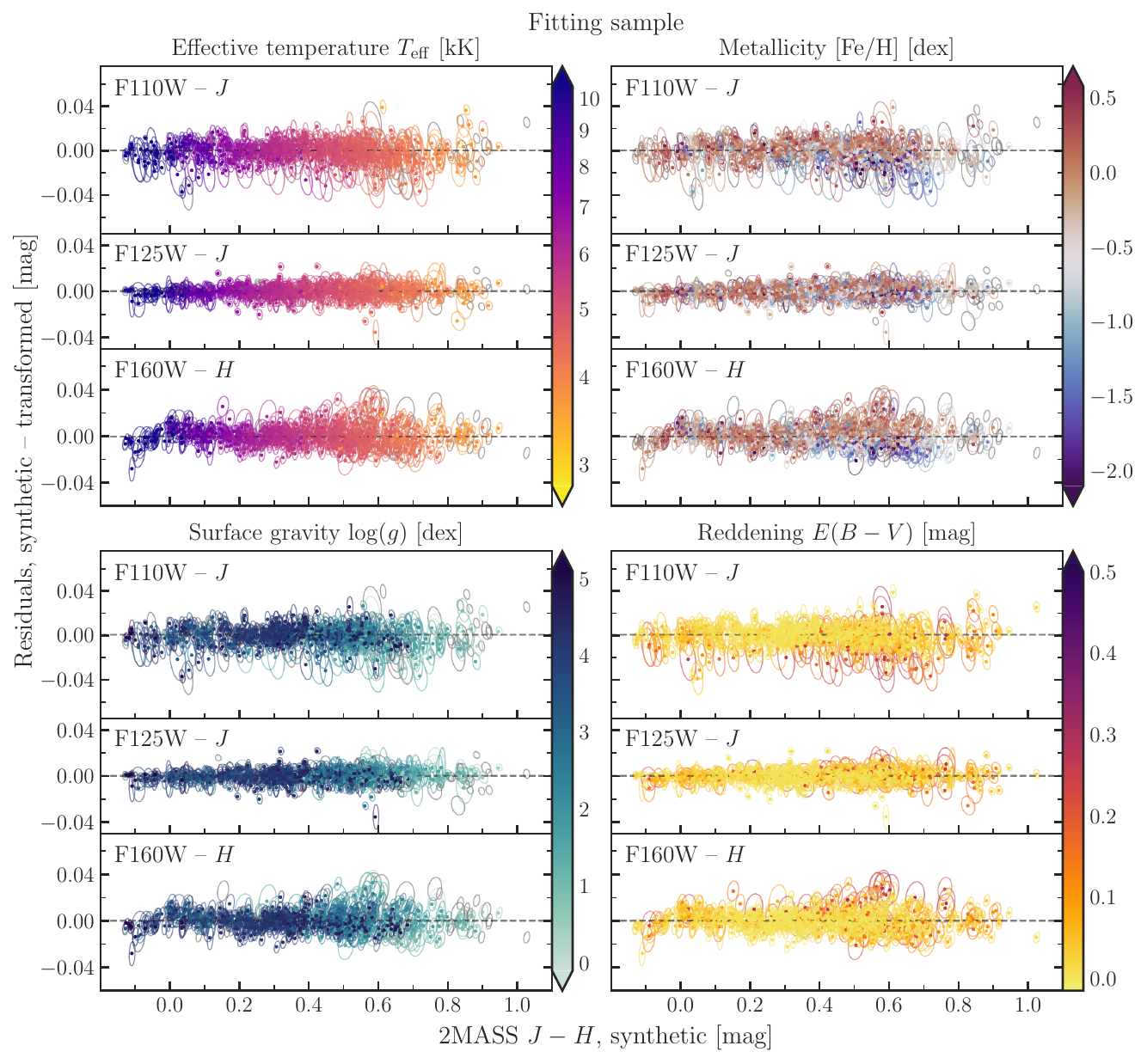}
    \caption{As \autoref{fig:jk_resid_fit} with respect to $J-H$ color on the $x$-axis.}
    \label{fig:jh_resid_fit}
\end{figure*}

\begin{figure*}
    \centering
    \includegraphics[width=0.9\textwidth]{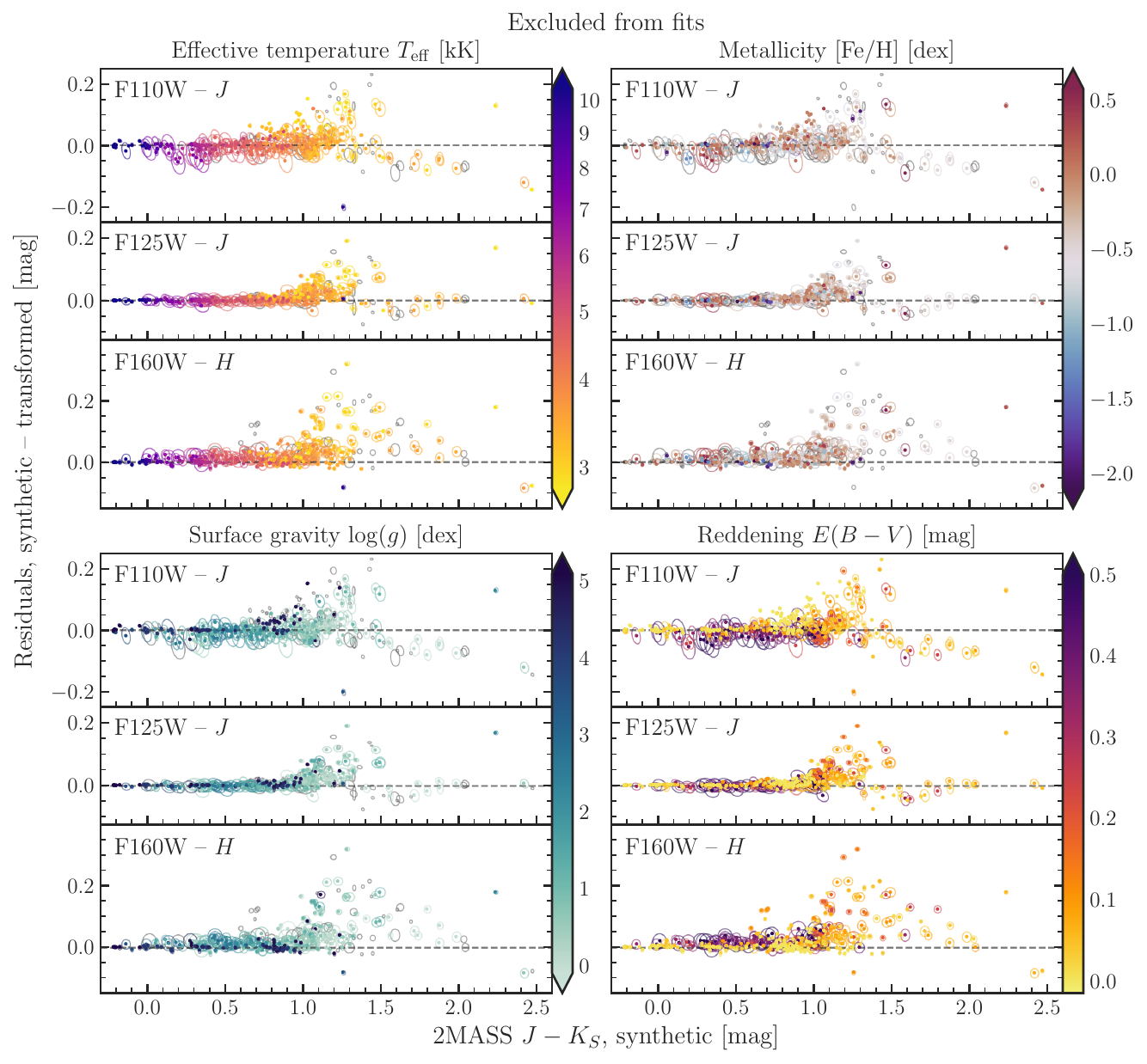}
    \caption{As \autoref{fig:jk_resid_fit}, but showing residuals for stars excluded from our fitting sample (424 spectra of 643 unique stars).
    Note the changes in both $x$- and $y$-axis ranges relative to figures~\ref{fig:jk_resid_fit} and \ref{fig:jh_resid_fit}.}
    \label{fig:jk_resid_cut}
\end{figure*}

\begin{figure*}
    \centering
    \includegraphics[width=0.9\textwidth]{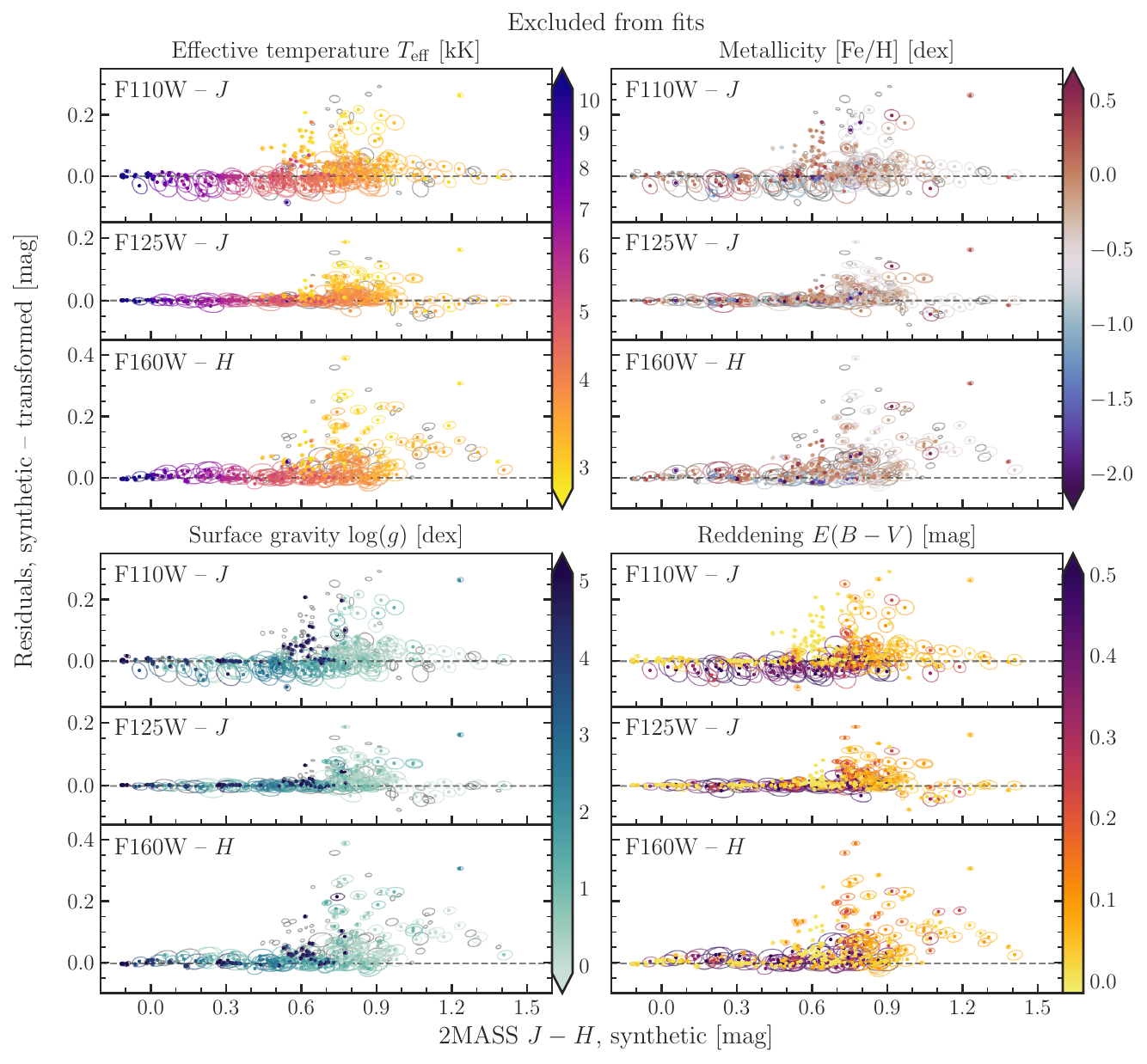}
    \caption{As \autoref{fig:jk_resid_cut}, with $J-H$ color on the $x$-axis.
    Again, note differences in $x$- and $y$-axis ranges between this and previous figures.}
    \label{fig:jh_resid_cut}
\end{figure*}

We now examine these transformations with respect to fundamental stellar properties.
In Figures~\ref{fig:jk_resid_fit} through \ref{fig:jh_resid_cut} we show residuals of the transformations color-coded by effective temperatures, surface gravities, and metallicities from the literature, as well as our adopted reddenings.
Figures~\ref{fig:jk_resid_fit} and \ref{fig:jh_resid_fit} show residuals of our fitting sample, and figures~\ref{fig:jk_resid_fit} and \ref{fig:jh_resid_fit} show residuals of stars excluded from the fits (i.e., those plotted in the right panel of \autoref{fig:fitting_selection}).
Stars without literature parameters are shown as open gray covariance ellipses.

Figures~\ref{fig:jk_resid_fit} and \ref{fig:jh_resid_fit} show that the fitting sample residuals are nearly all within 0.05 mag of our final filter conversions, and the majority are within 0.025 mag.
However, there may also be slight differences in the residuals that correlate with stellar parameters, leading to either increased scatter or small biases that vary systematically with intrinsic quantities, like metallicity and surface gravity, or observed quantities like extinction.
We note a possible metallicity dependence in the F110W--$J$ and F160W--$H$ residuals at $0.4 \lesssim J-H \lesssim 0.8$ (upper right of \autoref{fig:jh_resid_fit}), with metal-poor stars appearing to fall preferentially below the nominal fit (that is, slightly brighter in \HST filters relative to 2MASS than average), and metal-rich above (dimmer in \HST).
This is more or less expected, as higher metal content generally reddens stellar spectra and F110W and F160W both fall bluewards of $J$ and $H$ respectively.
However, we do not attempt to quantify this effect further given the heterogeneity of available metallicity information and possible confounding effects of surface gravity, dust, and/or specific abundance patterns such as alpha enhancement.

In the excluded sample (Figures~\ref{fig:jk_resid_cut} and \ref{fig:jh_resid_cut}), the most noticeable feature is the much higher level of scatter in the residuals overall, and the color dependence thereof.
In particular, the residual amplitude increases dramatically for red stars ($J-K_S \gtrsim 1$ or $J-H \gtrsim 0.5$) due to deep absorption features and variability (see \autoref{fig:distance_subtypes}).

\subsection{Stellar classifications} \label{ssec:dist_residuals}

\begin{figure*}[htb]
    \centering
    \includegraphics[width=\textwidth]{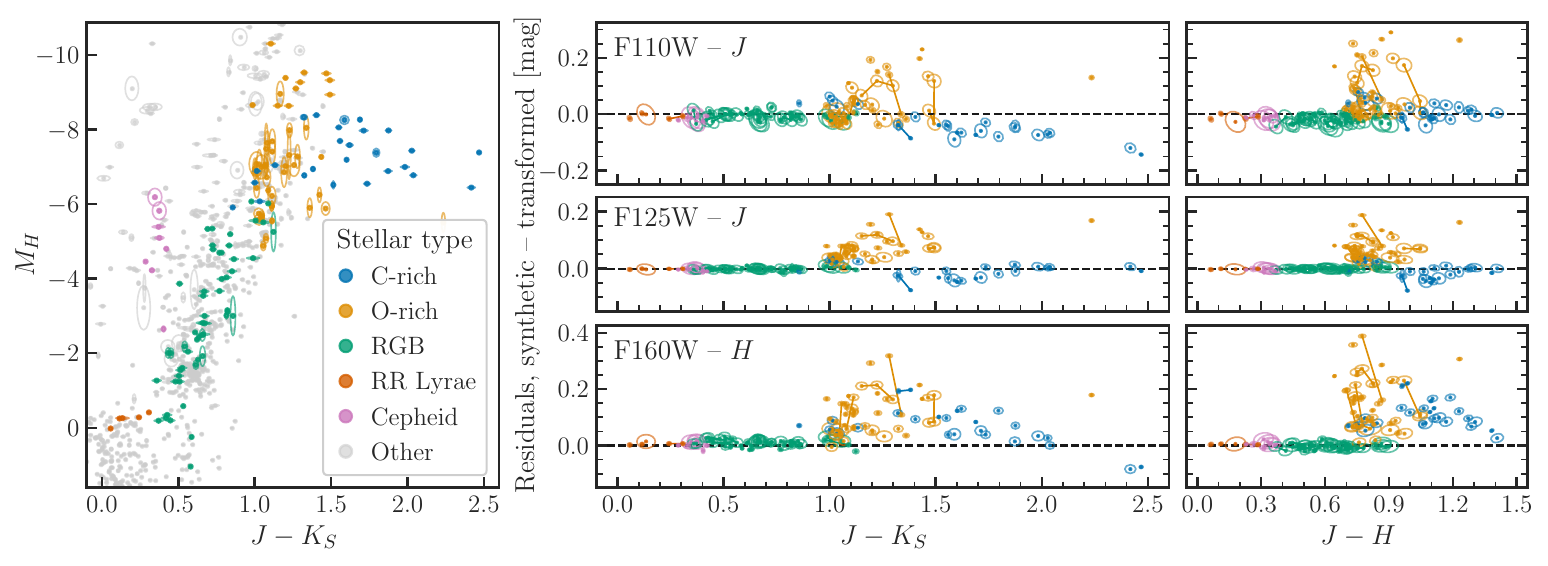}
    \caption{
    Synthetic 2MASS color-magnitude diagram (left) and filter transformation residuals (center, right) with several subtypes of evolved stars used as primary distance indicators highlighted by color.
    These include C- and O-rich TP-AGB variables in blue and orange respectively; first-ascent RGB stars in green; RR Lyrae variables in red; and classical Cepheids in pink.
    Repeat observations of the same stars are connected with solid lines.
  }
    \label{fig:distance_subtypes}
\end{figure*}

We now examine residuals for several subclasses of luminous evolved stars, many of which are commonly used in the literature as primary distance indicators.
\autoref{fig:distance_subtypes} compares synthetic photometry residuals for carbon-rich (blue) and oxygen-rich (orange) TP-AGB stars, RGB stars (green), RR Lyrae variables (RRL, red), and classical Cepheids (pink).
For stars with multiple library spectra taken at different phase points, we connect their synthetic magnitudes with solid lines to show variability effects. 

As seen for the reddest stars in the previous section, it is clear that TP-AGB stars show dramatically higher color-magnitude scatter relative to other stellar types due to temperature- and abundance-dependent absorption features.
Furthermore, the strength of these features---as well as the overall SED shape---can vary significantly in individual stars over the course of a thermal pulsation cycle; indeed, we see changes in band-to-band residuals of up to $\sim$0.2 mag in several of the O-rich variables.
We note that the XSL observations took simultaneous UV-optical-NIR spectra of such variables at multiple phase points to accurately model their net effects on the integrated light of simple stellar populations.
These results highlight the need for similar observing strategies---and careful choices of filters---when cross-calibrating TP-AGB magnitudes between near-infrared photometric systems.
This is especially salient in any distance-scale work that seeks to establish a common zeropoint between local calibrators best observed from the ground (e.g. Galactic stars with trigonometric parallaxes, the Magellanic Clouds) and space-based extragalactic observations of distant galaxies.

With the exception of the TP-AGB stars, however, we find that the transformations given in \autoref{tab:fitting} hold well for RR Lyrae, Cepheids, and RGB stars, and require no additional uncertainties beyond those reported in \autoref{tab:fitting}.

%%%%%%%%%%%%%%%%%%%%%%%%%%%%%%%%%%%%%%%%%%%%%%%%%%%%%%%%%%%%%%%%%%%%%%%%%%%%%%
\section{Conclusions} \label{sec:conc}
%%%%%%%%%%%%%%%%%%%%%%%%%%%%%%%%%%%%%%%%%%%%%%%%%%%%%%%%%%%%%%%%%%%%%%%%%%%%%%
We have derived new transformations between 2MASS and \HST broadband filters based on synthetic photometry of empirical stellar spectra.
While the majority of stars follow relations that are well-constrained as a function of broadband color, stars with $\teff \lesssim 3500$~K in particular diverge significantly from these nominal relations due to molecular absorption.
This is especially apparent in the case of both carbon- and oxygen-rich thermally pulsing AGB stars; furthermore, such stars' extreme and often irregular variability poses additional challenges for directly cross-calibrating them between systems.
While we cannot recommend a one-size-fits-all set of transformations for these stars, readers may find the individual synthetic magnitudes and literature parameters compiled in \autoref{app:tables} of use in this regard, depending on the application.

\subsection{Future Work}

We will use the data and methods presented here to predict similar transformations for a number of new and upcoming NIR photometric systems in a follow-up paper, including {\em{JWST}}/NIRCam, {\em{Roman}}/WFI, and {\em{Euclid}}/NISP.

While synthetic photometry is a well-tested and highly efficient method for predicting filter transformations, fully empirical calibrations are of course preferred when possible.
In a future paper series (Beaton et al.\ in prep), we will directly compare \HST WFC3/IR spectrophotometry of cool giants in the Magellanic Clouds to $JHK_S$ photometry of the same fields from the FourStar camera on Las Campanas Observatory's 6.5m Baade telescope.

\begin{acknowledgments}
We thank the anonymous referee for comments which improved and clarified this paper.
Support for this work was provided by NASA through HST program GO-15875. 
Support for this work was provided by the NSF through NSF grant AST-2108616.
Support for some of this work was provided by NASA through Hubble Fellowship grant \#51386.01 awarded to R.L.B. by the Space Telescope Science Institute, which is operated by the Association of  Universities for Research in Astronomy, Inc., for NASA, under contract NAS 5-26555.
We acknowledge the Undergraduate Summer Research Program (USRP) in the Princeton Department of Astrophysical Sciences that supported the initial work that formed this project; in particular, the efforts of Peter Melchior and Polly Strauss. An early verison of this work was presented as a Junior Paper to the Princeton Department of Physics and we thank Jo Dunkley and Lyman Page for their support as readers. 

%% SVO
This research has made use of the SVO Filter Profile Service (\url{http://svo2.cab.inta-csic.es/theory/fps/}) supported from the Spanish MINECO through grant AYA2017-84089.

%% 2MASS
This publication makes use of data products from the Two Micron All Sky Survey, which is a joint project of the University of Massachusetts and the Infrared Processing and Analysis Center/California Institute of Technology, funded by the National Aeronautics and Space Administration and the National Science Foundation.

This work has made use of data from the European Space Agency (ESA) mission \emph{Gaia} (\url{https://www.cosmos.esa.int/gaia}), processed by the \emph{Gaia} Data Processing and Analysis Consortium (DPAC, \url{https://www.cosmos.esa.int/web/gaia/dpac/consortium}). Funding for the DPAC has been provided by national institutions, in particular the institutions participating in the \emph{Gaia} Multilateral Agreement.

 This research has made use of NASA’s Astrophysics Data System.

This research has made use of the SIMBAD database, operated at CDS, Strasbourg, France. The original description of the SIMBAD service was published in \citet{simbad_2000}.

This research has made use of the VizieR catalogue access tool, CDS, Strasbourg, France (DOI: 10.26093/cds/vizier). The original description of the VizieR service was published in \citet{vizier2000}.

This research made use of the cross-match service provided by CDS, Strasbourg.

%%%
This research has used data, tools or materials developed as part of the EXPLORE project that has received funding from the European Union’s Horizon 2020 research and innovation programme under grant agreement No 101004214.

\end{acknowledgments}

%IRTF, XShooter, HST:STIS, HST:NICMOS, HST:WFC3/IR
\facilities{CDS, MAST, IRTF, XShooter, HST:STIS, HST:NICMOS, HST:WFC3/IR}

\software{Astropy \citep{2013A&A...558A..33A, 2018AJ....156..123A},
          Astroquery \citep{2017ascl.soft08004G, 2019AJ....157...98G},
          %BCcodes \citep{2014MNRAS.444..392C, 2018ascl.soft05022C},
          Dust\_extinction \citep{dust_extinction},
          Dustmaps \citep{2018JOSS....3..695M},
          Matplotlib \citep{2007CSE.....9...90H},
          NumPy \citep{numpy, 2020Natur.585..357H},
          Pandas \citep{pandas, mckinney2011},
          pwlf \citep{pwlf},
          PyVO \citep{2014ascl.soft02004G},
          Seaborn \citep{2021JOSS....6.3021W},
          SciPy \citep{2020NatMe..17..261V},
          Specutils \citep{specutils},
          % Scikit-learn \citep{sklearn},
          Stsynphot \citep{2020ascl.soft10003S},
          Synphot \citep{2018ascl.soft11001S}
}

\bibliography{bib,tmp,software}
\bibliographystyle{aasjournal}
\begin{appendix}

%%%%%%%%%%%%%%%%%%%%%%%%%%%%%%%%%%%%%%%%%%%%%%%%%%%%%%%%%%%%%%%%%%%%%%%%%%%%%%
\section{Machine-readable Tables} \label{app:tables}
%%%%%%%%%%%%%%%%%%%%%%%%%%%%%%%%%%%%%%%%%%%%%%%%%%%%%%%%%%%%%%%%%%%%%%%%%%%%%%

%%%%%%%%%%%%%%%%%%%%%%%%%%%%%%%%%%%%%%%%%%%%%%%%%%%%%%%%%%%%%%%%%%%%%
\autoref{tab:ids} lists all spectrum file names and their parent libraries, and the SIMBAD, 2MASS, and {\em{Gaia}} DR3 IDs of the target stars.
\autoref{tab:synphot} presents all synthetic 2MASS and \HST magnitudes and uncertainties used in this work. 
Band-to-band covariance information is available upon request.
\autoref{tab:dist_ebv} gives our adopted distance and reddening values.
In-text previews are limited to 10 lines to demonstrate the table structure and contents, and the full machine-readable tables are available in the online journal.

\begin{table*}[htbp]
  \centering
    \caption{Spectrum and star IDs} \label{tab:ids}
\begin{tabular}{llllll}
\hline \hline
Spectrum & Library & SIMBAD & 2MASS & {\em{Gaia}} DR3 \\
\hline
                 109vir\_stis\_003 & CALSPEC &                      * 109 Vir & 14461493+0153344 & 3655377057091634304 \\
                  10lac\_stis\_007 & CALSPEC &                      *  10 Lac & 22391567+3903011 & 1908095850396090880 \\
                 16cygb\_stis\_003 & CALSPEC &                    *  16 Cyg B & 19415198+5031032 & 2135550755683407232 \\
             1732526\_stisnic\_007 & CALSPEC &                TYC 4424-1286-1 & 17325264+7104431 & 1651137131123978112 \\
             1740346\_stisnic\_005 & CALSPEC &                 TYC 4207-219-1 & 17403468+6527148 & 1633143932573832448 \\
             1743045\_stisnic\_007 & CALSPEC &        2MASS J17430448+6655015 & 17430448+6655015 & 1634280312200704768 \\
             1757132\_stiswfc\_004 & CALSPEC &                 TYC 4212-455-1 & 17571324+6703409 & 1633585107317144960 \\
          1802271\_stiswfcnic\_004 & CALSPEC &              BPS BS 17447-0067 & 18022716+6043356 & 2158745262705810304 \\
             1805292\_stisnic\_006 & CALSPEC &                TYC 4209-1396-1 & 18052927+6427520 & 2161093682102883712 \\
             1808347\_stiswfc\_004 & CALSPEC &                TYC 4433-1800-1 & 18083474+6927286 & 2260019315938461952 \\
\hline \hline
\end{tabular}
\end{table*}

\begin{table*}[htbp]
  \centering
    \caption{Synthetic magnitudes} \label{tab:synphot}
\begin{tabular}{l|rrrrrrrrrrrr}
\hline \hline
{} & \multicolumn{2}{c}{$J$} & \multicolumn{2}{c}{$H$}  & \multicolumn{2}{c}{$K_S$} & \multicolumn{2}{c}{F110W} & \multicolumn{2}{c}{F125W} & \multicolumn{2}{c}{F160W} \\ 
Spectrum &       mag &     err &       mag &     err &  mag &     err &  mag &     err & mag &     err & mag &     err  \\ \hline
109vir\_stis\_003                  &  3.675 & 0.008 &  3.668 & 0.008 &  3.673 & 0.008 &  3.667 & 0.007 &  3.675 & 0.008 &  3.672 & 0.008 \\
10lac\_stis\_007                   &  5.273 & 0.013 &  5.388 & 0.009 &  5.486 & 0.007 &  5.212 & 0.015 &  5.273 & 0.013 &  5.356 & 0.010 \\
16cygb\_stis\_003                  &  5.017 & 0.008 &  4.676 & 0.008 &  4.635 & 0.008 &  5.140 & 0.007 &  5.012 & 0.008 &  4.732 & 0.008 \\
1732526\_stisnic\_007              & 12.264 & 0.016 & 12.216 & 0.013 & 12.210 & 0.012 & 12.287 & 0.017 & 12.263 & 0.016 & 12.220 & 0.014 \\
1740346\_stisnic\_005              & 12.059 & 0.016 & 11.977 & 0.013 & 11.967 & 0.012 & 12.101 & 0.016 & 12.059 & 0.016 & 11.990 & 0.014 \\
1743045\_stisnic\_007              & 12.933 & 0.019 & 12.820 & 0.015 & 12.805 & 0.013 & 12.990 & 0.020 & 12.933 & 0.019 & 12.838 & 0.016 \\
1757132\_stiswfc\_004              & 11.262 & 0.012 & 11.159 & 0.009 & 11.156 & 0.009 & 11.305 & 0.014 & 11.261 & 0.012 & 11.176 & 0.010 \\
1802271\_stiswfcnic\_004           & 11.827 & 0.013 & 11.811 & 0.011 & 11.803 & 0.012 & 11.825 & 0.016 & 11.825 & 0.014 & 11.819 & 0.011 \\
1805292\_stisnic\_006              & 12.017 & 0.016 & 11.973 & 0.013 & 11.972 & 0.012 & 12.037 & 0.017 & 12.017 & 0.016 & 11.978 & 0.014 \\
1808347\_stiswfc\_004              & 11.623 & 0.014 & 11.539 & 0.011 & 11.550 & 0.009 & 11.665 & 0.016 & 11.622 & 0.014 & 11.554 & 0.011 \\
\hline \hline
\end{tabular}
\end{table*}

\begin{table*}[htbp]
  \centering
    \caption{Adopted distances and reddenings} \label{tab:dist_ebv}
\begin{tabular}{l|rrrcccc}
\hline \hline
{} & \multicolumn{1}{c}{Distance $D$} & \multicolumn{1}{c}{$\sigma_D^{\mathrm{low}}$} & \multicolumn{1}{c}{$\sigma_D^{\mathrm{high}}$} & \multicolumn{1}{c}{$E({B-V})$} & \multicolumn{1}{c}{$\sigma_{E(B-V)}$} & $D$ ref & $E$ ref   \\
SIMBAD & \multicolumn{1}{c}{pc} & \multicolumn{1}{c}{pc} & \multicolumn{1}{c}{pc} & \multicolumn{1}{c}{mag} & \multicolumn{1}{c}{mag} & {} & {}   \\
\hline
                     * 109 Vir &    41.1550 &    -0.3615 &    0.4828 & 0.0064 & 0.0004 &      1 &       13 \\
                     *  10 Lac &   454.9370 &   -32.5998 &   35.1707 & 0.0801 & 0.0036 &      1 &       13 \\
                   *  16 Cyg B &    21.1187 &    -0.0085 &    0.0081 & 0.0035 & 0.0003 &      1 &       13 \\
               TYC 4424-1286-1 &  1170.3805 &   -22.7738 &   24.0533 & 0.0196 & 0.0116 &      1 &       14 \\
                TYC 4207-219-1 &  1240.0450 &   -30.6903 &   28.6683 & 0.0287 & 0.0099 &      1 &       14 \\
       2MASS J17430448+6655015 &  1771.9917 &   -35.9045 &   37.0383 & 0.0363 & 0.0147 &      1 &       14 \\
                TYC 4212-455-1 &  1002.1997 &   -33.1524 &   38.7034 & 0.0508 & 0.0079 &      1 &       14 \\
             BPS BS 17447-0067 &  1514.9535 &   -52.2798 &   41.7642 & 0.0520 & 0.0102 &      1 &       14 \\
               TYC 4209-1396-1 &  1433.7218 &   -39.9004 &   33.1952 & 0.0316 & 0.0106 &      1 &       14 \\
               TYC 4433-1800-1 &   918.7812 &   -15.7393 &   13.4752 & 0.0415 & 0.0122 &      1 &       14 \\
\hline \hline
\end{tabular}

\tablerefs{1: {\citet{2021AJ....161..147B}}; 2: {\citet{2007A&A...474..653V}}; 3: {\citet{2018AJ....156...58B}}; 4: {\citet{2022A&A...657A.131M}}; 5: {\citet{2021MNRAS.505.5957B}}; 6: {\citet{2020A&A...633A..99C}}; 7: {\citet{2010AJ....139.1808S}}; 8: {\citet{2020MNRAS.493..468D}}; 9: {\citet{2010A&ARv..18...67T}}; 10: {\citet{2007AJ....134.2200S}}; 11: {\citet{2020ApJ...904...13G}}; 12: {\citet{2019Natur.567..200P}}; 13: {\citet{2022AA...664A.174V}}; 14: {\citet{2022arXiv220606710D}}; 15: {\citet{2021ApJS..252...23S}}
}

\end{table*}

%%%%%%%%%%%%%%%%%%%%%%%%%%%%%%%%%%%%%%%%%%%%%%%%%%%%%%%%%%%%%%%%%%%%%

%%%%%%%%%%%%%%%%%%%%%%%%%%%%%%%%%%%%%%%%%%%%%%%%%%%%%%%%%%%%%%%%%%%%%%%%%%%%%%
\section{Comparison of Extinction Maps} \label{app:extinction}
%%%%%%%%%%%%%%%%%%%%%%%%%%%%%%%%%%%%%%%%%%%%%%%%%%%%%%%%%%%%%%%%%%%%%%%%%%%%%
We investigated a number of available line-of-sight extinction estimates for our sample of library stars, both from large-scale extinction maps and from parameter fits to individual stars. 
Of these, 
    five are from three-dimensional (3D) maps derived from aggregate spectrophotometric and parallax information \citep{2022AA...664A.174V, 2022AA...661A.147L, 2018AA...616A.132L, 2019ApJ...887...93G, 2019MNRAS.483.4277C}; 
    three are from SED fitting techniques applied to individual library stars, including the values adopted for or derived from the original spectral library data \citep{2022AA...658A..91A, 2022arXiv220606138A}; 
    and the remaining three are from two-dimensional (2D) total line-of-sight maps \citep{2022arXiv220606710D, 2021ApJS..252...23S, 2011ApJ...737..103S}.

A summary of all extinction data sources we considered is given in \autoref{tab:extinction}, including primary references, applied conversion factors from original data units to $E(B-V)$ if needed, and numbers of library stars for which extinction data are a) available for a given star or distance-independent line of sight, and b) of those how many are considered valid measurements.
For 3D maps we follow the \texttt{dustmaps} convention, where $E(B-V)$ values are considered valid for a given sightline within a specified range of distances.
We consider single-star $E(B-V)$ measurements valid if they are free from data or fitting quality flags as defined by respective references, if available.

\begin{figure}[htbp]
    \centering
    \includegraphics[width=\textwidth]{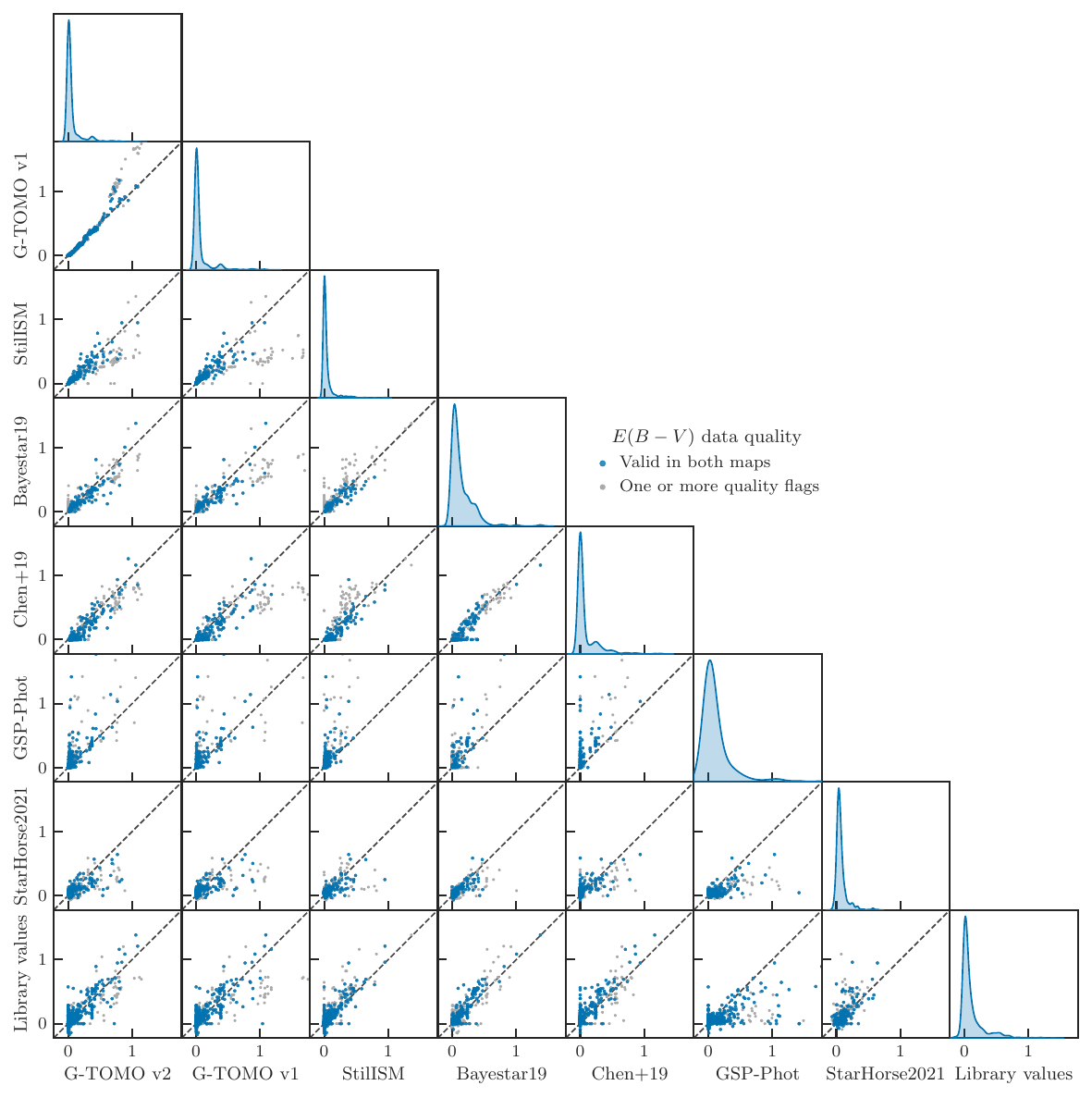}
    \caption{
    Pair plots of $E(B-V)$ measurements for stars common to different extinction catalogs. Stars with high-quality measurements in both $E(B-V)$ maps per pair are shown by blue points, and stars with data quality flags in one or both maps are shown in light gray.
    Summary data for individual tables are given in \autoref{tab:extinction}, and median absolute deviations of pairwise differences in \autoref{tab:extinction_comparison}. 
    }
    \label{fig:ebv_pairplot}
\end{figure}

Figure~\ref{fig:ebv_pairplot} shows pairwise comparisons for $E(B-V)$ values from the 3D maps and individual stellar fits, and median absolute deviations of each of these are given in \autoref{tab:extinction_comparison}.
The majority of $E(B-V)$ measurements agree to within 0.05 or 0.1 mag, with the exception of GSP-Phot, which shows deviations of up to 0.25 mag.

%%%%%%%%%%%%%%%%%%%%%%%%%%%%%%%%%%%%%%%%%%%%%%%%%%%%%%%%%%%%%%%%%%%%%
\begin{table*}[htbp]
    \centering
    \caption{Summary of extinction data sources}
    \label{tab:extinction}
    \begin{tabular}{c | c c c c l l}
    \hline
     Kind & Name   & $N^{\star}_{\rm{total}}$ & $N^{\star}_{\rm{valid}}$ & $E(B-V)$ & Reference(s) & Comments \\
     {} & {} & {} & {} & conversion & {} & {} \\
    \hline \hline
    % \multicolumn{6}{c}{3D line-of-sight maps} \\ \hline
    3D maps &       G-TOMO v2 &      980 &      808 & $3.1^{-1}$ &  \citealt{2022AA...664A.174V} & {}                          \\
         {} &        G-TOMO v1 &      980 &      808 & $3.1^{-1}$ &  \citealt{2022AA...661A.147L} & {}                       \\
         {} &          StilISM &      939 &      724 &        --- &  \citealt{2018AA...616A.132L} & {}                       \\
         {} &       Bayestar19 &      891 &      238 &      0.884 & \citealt{2019ApJ...887...93G} & {}                       \\
         {} &          Chen+19 &      480 &      377 &       0.75 & \citealt{2019MNRAS.483.4277C} & \emph{Gaia} DR2 passbands \\
\hline
Single-star &   StarHorse2021 &      619 &      425 & $3.3^{-1}$ &  \citealt{2022AA...658A..91A} & {}                       \\
         {} &         GSP-Phot &      533 &      533 & $3.1^{-1}$ & \citealt{2022arXiv220606138A} & {}                       \\
         {} &   Library values &      926 &      926 &        --- & \citealt{2009ApJS..185..289R}, & {}                       \\
         {} &   {} &      {} &      {} &        {} & \citealt{2017ApJS..230...23V}, & {}                         \\
         {} &   {} &      {} &      {} &        {} & \citealt{2022AA...660A..34V} & {}                         \\
\hline
    2D maps & \emph{Gaia} TGE &     1020 &     1020 & $3.1^{-1}$ & \citealt{2022arXiv220606710D} & {}                       \\
         {} &              SFD &     1061 &     1061 &        --- & \citealt{2011ApJ...737..103S} & {}                       \\
         {} &             OGLE &       76 &       76 &      0.808 & \citealt{2021ApJS..252...23S} &    Magellanic Clouds only \\
    \hline \hline
    \end{tabular}
\end{table*}

\begin{table}[ht]
\centering
\caption{Median absolute deviations of pairwise $E(B-V)$ differences for 3D maps and single-star fits}
\label{tab:extinction_comparison}
\begin{tabular}{l@{}rrrrrrrr}
\hline
 & G-TOMO v2 & G-TOMO v1 & StilISM & Bayestar19 & Chen+19 & StarHorse2021 & GSP-Phot & Library \\
\hline \hline
G-TOMO v2     &        --- &      0.010 &    0.014 &       0.070 &    0.056 &          0.055 &     0.111 &    0.049 \\
G-TOMO v1     &      0.010 &        --- &    0.016 &       0.086 &    0.066 &          0.058 &     0.110 &    0.050 \\
StilISM       &      0.014 &      0.016 &      --- &       0.059 &    0.045 &          0.045 &     0.087 &    0.042 \\
Bayestar19    &      0.070 &      0.086 &    0.059 &         --- &    0.053 &          0.050 &     0.243 &    0.072 \\
Chen+19       &      0.056 &      0.066 &    0.045 &       0.053 &      --- &          0.064 &     0.250 &    0.076 \\
StarHorse2021 &      0.055 &      0.058 &    0.045 &       0.050 &    0.064 &            --- &     0.105 &    0.056 \\
GSP-Phot      &      0.111 &      0.110 &    0.087 &       0.243 &    0.250 &          0.105 &       --- &    0.169 \\
Library       &      0.049 &      0.050 &    0.042 &       0.072 &    0.076 &          0.056 &     0.169 &      --- \\
\hline \hline
\end{tabular}
\end{table}

%%%%%%%%%%%%%%%%%%%%%%%%%%%%%%%%%%%%%%%%%%%%%%%%%%%%%%%%%%%%%%%%%%%%%

%%%%%%%%%%%%%%%%%%%%%%%%%%%%%%%%%%%%%%%%%%%%%%%%%%%%%%%%%%%%%%%%%%%%%
%%%%%%%%%%%%%%%%%%%%%%%%%%%%%%%%%%%%%%%%%%%%%%%%%%%%%%%%%%%%%%%%%%%%%

\end{appendix}

\end{document}